\documentclass[pra,showpacs,twocolumn]{revtex4-1}
\usepackage{bm,graphicx,amsmath, color}

\newcommand{\beq}{\begin{equation}}
\newcommand{\enq}{\end{equation}}

\newcommand{\veci}{\bf i}
\newcommand{\vecj}{\bf j}

\begin{document}
\title{Confined $p$-band Bose-Einstein condensates}
\author{Fernanda Pinheiro$^{1,2}$}
\email{fep@fysik.su.se}
\author{Jani-Petri Martikainen$^{2,3}$}
\author{Jonas Larson$^{1,4}$}
\address{$^1$Department of Physics, Stockholm University, 
SE-106 91 Stockholm, Sweden}
\address{$^2$NORDITA, SE-106 91 Stockholm, Sweden}
\address{$^3$Aalto University, P.O. Box 1510, FI-00076 Aalto, Finland}
\address{$^4$Institut f\"ur Theoretische Physik, Universit\"at zu K\"oln, K\"oln, De-50937, Germany}
\date{\today}

\begin{abstract}
We study bosonic atoms on the $p$-band of a two dimensional optical
square lattice in the presence of a confining trapping
potential. Using a mean-field approach, we show how the anisotropic
tunneling for $p$-band particles affects the cloud of condensed atoms
by characterizing the ground state density and the coherence
properties of the atomic states both between sites and atomic
flavors. 
In contrast to the usual results based on the LDA, the atomic density can become anisotropic. 
This anisotropic effect is especially
pronounced in the limit of weak atom-atom interactions and of weak
lattice amplitudes, i.e. when the properties of the ground state are
mainly driven by the kinetic energies. We also investigate how the
trap influences known properties of the non-trapped case. In
particular, we focus on the behavior of the anti-ferromagnetic
vortex-antivortex order, which for the confined system, is shown to
disappear at the edges of the condensed cloud.
\end{abstract}


\maketitle

\section{Introduction}
With refined experimental techniques in trapping and cooling, atomic
gases have become prime candidates for studies of mesoscopic quantum
phenomena~\cite{blochreview}. Among different possible
experimental configurations ~\cite{blochreview,maciekreview}, systems
of cold atoms subjected to optical lattices constitute one of the
most active topics of the current research in the field. In the
ultracold limit, these setups may serve as  {\it quantum simulators}
which can be used to test actual models of condensed matter theories
in a precise way~\cite{maciekreview}. In fact, the degree of
experimental control in optical lattice systems is so great, that by
tuning the parameters of the lattice the atoms can be moved into
the strongly correlated regime, therefore allowing
for the study of a variety of phenomena which include quantum
phase transitions~\cite{mottsf}. Beyond experimental
manipulations of the ground state, the versatility of these systems
also makes it possible to experimentally prepare certain excited states. In
this respect, of particular interest are the states of bosons
restricted to the first excited energy bands of the
lattice, the so called $p$-band bosons.

Qualitatively, the physics of $p$-band bosons is considerably
different from the well studied systems where the bosons are only 
restricted to the lowest band ($s$-band bosons). The reason for this
can be intuitively understood from the isotropic square and cubic
lattices, where the symmetry of the lattice implies a double
(square lattice) and triple (cubic lattice) degeneracy~\cite{plew, isaksson}
on the $p$-band. In solid state systems such degeneracies
could be removed via Jahn-Teller effects, but since here the
lattice is imposed from the outside, the degeneracy is robust.
This degeneracy motivates the description of the atomic states
in terms of orbitals related to the corresponding localized Wannier
functions, characterized by a node in each of the spatial directions. 
In the direction of the node, the Wannier functions are also 
broader and since this directly influences the ease of tunneling between sites, it
directly affects the dynamical properties of the system. Since the
properties of the tunneling of $p$-band bosons are dramatically
altered from the ones on the $s$-band, a rich variety of
novel quantum phases~\cite{ph1,ph2,ph3,ph4} can appear. When interactions 
are taken into account, it has also been argued that 
in the limit of very strong atom-atom interactions, 
atomic population can move to higher energy bands, affecting thus the
expected ground state properties of ultracold 
atoms in optical lattices~\cite{pinf1,pinf2,pinf3,pinf4, pinf5,
  pinftrap}. The broadening of the onsite wave-functions, for example,
was experimentally verified via microwave
spectroscopy~\cite{blochmicro}. In addition, signatures of 
(strong) interaction induced higher bands physics could also be seen in
non-equilibrium configurations, through the
mapping of collapse-revivals structures in the atomic 
density~\cite{blochcr} (see also Ref.~\cite{crteo}).  Surprising effects
are also present in the limit of weak interactions. In fact, it was
recently observed~\cite{hemmerich1,hemmerich2} that due to unusual
dispersions, the physics of $p$-band bosons appears responsible for
unconventional condensation, where non-zero momentum states~\cite{nzp}
are occupied. We should point out, however, that even though experiments
concerning $p$-band physics have been restricted to one dimensional,
square or cubic lattices~\cite{hemmerich1,hemmerich2,pexp1,pexp2}, several
theoretical predictions have been made for other lattice
configurations~\cite{pother}.

In experiments, optical lattice systems are 
generally subjected to an external confining trap. Although
it is known that even for $s$-band bosons, the presence of the trap can add
important features to the physics of the system~\cite{wedding}, 
all the aforementioned theoretical studies of
$p$-band bosonic systems neglect effects originating from the
confining trap potential. Thus, it is important to study how the
inclusion of a trap affects the $p$-band physics. 
For example, in the case of a two dimensional (2D) lattice
it is characteristic of $p$-band bosons to have
tunneling coefficients with different amplitudes in different
directions. In the non-trapped case, this property of anisotropic 
tunneling together with the
properties of homogeneous density distributions yields a corresponding
ground state which has an anti-ferromagnetic order with
vortex/anti-vortex states on every second site (also known as the
state of staggered orbital angular momentum)~\cite{isaksson,ph4}.
In trapped systems, however, the property of anisotropic tunneling 
necessarily introduces density inhomogeneities
which break the population balance between
different possible atomic states (here corresponding to the two
possible orbitals of the 2D lattice). This also gives rise to
physics beyond the one captured by using the local density approximation (LDA). The fate 
of the anti-ferromagnetic order in the presence of the trap
is then unclear.  

In this paper we study this issue and address also other effects and
properties which arise when $p$-band bosons are confined by an
external potential. We mostly restrict the analysis to 2D,
but discuss how the obtained results generalize to 3D. 
The analysis is based on the ideal gas theory and
a mean-field approach, where we assume the system
to be deep in the region of the superfluid phase. We start
by presenting the theoretical framework and follow with the study of the
ideal system at finite temperatures, where the critical temperature
for condensation in a non-interacting $p$-band bosonic gas is calculated. 
We then show that for a symmetric square lattice, the zero temperature
order parameter of the condensed ground state is
complex also in the presence of a trap, but the vortex/anti-vortex
structure can be lost. In particular, the ground state for the
$p$-band atomic densities of the two flavors are shown to be different
except for when the system is driven
into the Thomas-Fermi (TF) regime in which case we can neglect effects
stemming from the kinetic tunneling energy. We complete the study with an
analysis of the zero temperature properties of an asymmetric
lattice. We find that due to splitting of the
$p$-band degeneracy, the ground state properties may be 
sensitive to small changes in the two lattice amplitudes.  

It is important to point out that our analysis
is carried out when influence from other bands have been omitted. 
The validity of this assumption is specially tested in the harmonic
approximation, where two $p$-band atoms become degenerate with one
$s$- and one $d$- atom. In fact, due to a 'reduced final density of
states for scattering processes' \cite{pexp2}, these decays can
be significantly suppressed \cite{isaksson} and the lifetimes of the atoms in
$p$- orbitals become 1-2 order of magnitude larger than typical
tunneling times. In addition, outside the harmonic approximation as
the case considered throughout this paper, the actual anharmonicity of
the lattice breaks the ($p + p \rightarrow s + d$) degeneracy for
almost all quasi momenta, suppressing further such loss processes.

\section{Derivation of the effective model Hamiltonian}

\subsection{Hamiltonian for $p$-band bosons}
In terms of the field operators $\hat{\Psi}(\vec{r}')$, the dynamics
of the weakly interacting Bose gas can be described by the Hamiltonian 
\begin{equation}\label{Field_Ham}
\begin{array}{lll}
H & = &\displaystyle{ \int d\vec{r}'\Big\{\hat{\Psi}^{\dagger}(\vec{r}')\left[-\frac{\hbar^2\nabla^2}{2m} +
V(\vec{r}')\right]\hat{\Psi}(\vec{r}')}\\ \\
& & +\displaystyle{\frac{\tilde{U}_0}{2}\hat{\Psi}^{\dagger}(\vec{r}')\hat{\Psi}^{\dagger}(\vec{r}')\hat{\Psi}(\vec{r}')\hat{\Psi}(\vec{r}')}\Big\},
\end{array}
\end{equation}
where $m$ corresponds to the mass of the particles, $\tilde{U}_0$ to the
strength of the interparticle interaction, and $V(\vec{r}')$ accounts for
the effects of external potentials acting on the system. The field operators $\hat{\Psi}(\vec{r}')$ and $\hat{\Psi}^\dagger(\vec{r}')$ annihilate and create a particle at position $\vec{r}'$ respectively, and obey the standard boson commutation relation $\left[\hat{\Psi}(\vec{r}''),\hat{\Psi}^\dagger(\vec{r}')\right]=\delta(\vec{r}''-\vec{r}')$. In this work
we consider a trapped system in 2D with $V(\vec{r}') =
V_{latt}(\vec{r}') + V_{trap}(\vec{r}')$, where the optical lattice potential
\begin{equation}
V_{latt}(\vec{r}') = \tilde{V}_{x}\sin^2\left(kx'\right) +
\tilde{V}_{y}\sin^2\left(ky'\right) 
\end{equation}
has amplitudes and wave vector given, respectively, by
$\tilde{V}_{\alpha}$, $\alpha \in \{x, y\}$, and $k=2\pi/\lambda$,
with $\lambda$ being the wave length of the applied lasers, and where
\begin{equation}
V_{trap}(\vec{r}') = \frac{m\tilde{\omega}^2}{2}\left(x'^2 + y'^2\right)
\end{equation}
describes the action of an overall slowly varying harmonic trap with
frequency $\tilde{\omega}$. 

The common practice in the study of many-body systems subjected to
periodic potentials consists in the expansion of the many-body
Hamiltonian in terms of a suitable basis, generally constructed from
its corresponding non-interacting part. In fact, the invariance under
discrete translations of the lattice implies conservation of
quasi-momentum and an energy spectrum having a band structure,
which therefore immediately suggest the use of Bloch functions. Here,
however, the presence of the trap breaks translational invariance and
implies a finite size for the system, consequently destroying the
symmetries that rigorously justify theoretical treatment in these
terms. On the other hand, the smoothness of the potential implies 
that its characteristic length
scale fulfills the condition 
$l_{trap}=\sqrt{\hbar/m\tilde{\omega}}\gg\lambda/2$, and thus we can
implement the effects of the trap in each site, by only shifting 
the onsite energies and assuming that the onsite orbitals remain the same
in the absence of a trap. This means that locally the system is
still effectively periodic, and that a satisfactory approximation can be obtained 
from the traditional framework.  

Before carrying out the expansion of the field operators we define
dimensionless parameters by taking the recoil energy
$E_r=\hbar^2k^2/2m$ as the energy scale (i.e. all 
energies are scaled by this quantity) and the inverse wave vector as
the typical length scale $l=\lambda/2\pi$, which produces a
dimensionless trap frequency given by $\omega=\sqrt{2}m\tilde{\omega}/\hbar
k^2$. In these terms, the trapping potential becomes
$V(\vec{r})=\omega^2\left(x^2+y^2\right)/2$,  with 
$x=kx'$ and $y=ky'$ the dimensionless positions. From now on,
we assume these units in all the derivations so that
resulting equations are dimensionless. 
As a first step, we construct the bosonic operators $\hat{b}_{\nu\bf{q}}$ and
$\hat{b}_{\nu\bf{q}}^\dagger$ which create and annihilate,
respectively, one particle delocalized in the Bloch state $\phi_{\nu \bf{q}}(\vec{r})$ 
of quasi-momentum ${\bf q}=(q_x,q_y)$ in the
$\nu$-th energy band, and use it to write 
\begin{equation}
\begin{array}{l}
\displaystyle{\hat{\Psi}^{\dagger}(\vec{r}) = \sum_{\nu \bf{q}}\phi^{*}_{\nu \bf{q}}(\vec{r})\,\hat{b}^{\dagger}_{\nu \bf{q}}},\\ \\
\displaystyle{ \hat{\Psi}(\vec{r}) = \sum_{\nu \bf{q}}\phi_{\nu \bf{q}}(\vec{r})\,\hat{b}_{\nu \bf{q}}},
\end{array}
\end{equation}
where the $\nu$-sum runs over all energy bands, and the ${\bf q}$-sum is
over the first Brillouin zone. We also use the above expressions to 
construct the site-localized Wannier functions, where the operators read
\begin{equation}\label{wanexp}
\begin{array}{l}
\displaystyle{\hat{\Psi}^{\dagger}(\vec{r}) = \sum_{\nu j}w^*_{\nu
    {\bf R}_{\bf j}}(\vec{r})\,\hat{a}^{\dagger}_{\nu{\bf j}}},\\ \\
\displaystyle{ \hat{\Psi}(\vec{r}) = \sum_{\nu \bf{j}}w_{\nu {\bf R}_{\bf j}}(\vec{r})\,\hat{a}_{\nu{\bf j}}}.
\end{array}
\end{equation}
Here, ${\bf R}_{\bf j} = (x_{\bf j},y_{\bf j})=(\pi j_x, \pi j_y)$ labels the coordinates of the $j$'th site of
the lattice (${\bf j}=(j_x,j_y),\,j_x,j_y\in\mathcal{N}$), and
$\hat{a}_{\nu\bf{j}}$ ($\hat{a}_{\nu\bf{j}}^\dagger$) annihilate
(create) a particle in the Wannier state $w_{\nu {\bf R}_{\bf
    j}}(\vec{r})$. For completeness, the relation between Wannier and
Bloch functions is given by
\begin{equation}
w_{\nu{\bf R}_{\bf j}}(\vec{r}) = \sum_{\bf q} e^{-i{\bf q}\cdot{\bf R}_{\bf j}}\phi_{\nu {\bf q}}(\vec{r}).
\end{equation}
As a second step in deriving an effective model described by the
Hamiltonian of Eq.~(\ref{Field_Ham}), we choose the expansion of the
many-body Hamiltonian in terms of (\ref{wanexp}) and introduce some
approximations. Our option for this picture
is justified by the fact that while considerably simpler for
the practical implementations, the use of
Wannier basis together with the tight-binding approximation can
still provide a good description as long as the lattice is
deep enough~\cite{jonasNJP}. In
addition to restricting the hopping to nearest-neighbors (tight-binding), we
truncate the expansion of the field operators to include only the
$p$-bands.

As the last step of our derivation, we clarify the used
terminology. For a square lattice, the two $p$-band Wannier functions
at each site ${\bf j}$ are characterized by a node along either the $x$- or
$y$- directions. Therefore we call atoms with orbital wavefunctions, $w_{x{\bf j}}(\vec{r})$ and
$w_{y{\bf j}}(\vec{r})$, respectively as $x$- and $y$-flavors~\cite{isaksson}, and for completeness we give their explicit
expressions 
\begin{equation}\label{flavor_wfx}
\begin{array}{l}
w_{x{\bf j}}( \vec{r}) = w_{2j_x}(x)w_{1j_y}(y),\\ \\
w_{y{\bf j}}( \vec{r}) = w_{1j_x}(x)w_{2j_y}(y).\\ \\
\end{array}
\end{equation}
From this, the nature of the node-structure becomes clear. It is a
direct consequence of the nodal structure of the Wannier functions
$w_{2j}(x)$ and  $w_{1j}(x)$. An $x$-flavor (or equivalently
$p_x$-orbital) atom, thus, not only has a wavefunction with a node along the
$x$-direction, but also a broader distribution along $x$. Accordingly,
the opposite is true for atoms in the $y$-flavor.  This property
directly affects the tunneling properties of the atoms in this system.

Putting everything together, we can 
write down the resulting many-body Hamiltonian 
\begin{equation}\label{manyham}
H = H_0 + H_{nn} + H_{FD},
\end{equation}
with the ideal part given by
\begin{equation}
H_0 = -\sum_{\alpha,\beta}\sum_{\langle{\bf i}{\bf j}\rangle_\alpha} 
t_{\alpha\beta}\,\hat{a}_{\beta{\bf i}}^{\dagger}\hat{a}_{\beta{\bf
j}} + \sum_\alpha \sum_{\bf j}V_{trap}({\bf R}_{\bf j})\hat{n}_{\alpha{\bf j}},
\end{equation}
where $\sum_{\langle\veci\vecj\rangle_\alpha}$ refers to the sum over nearest
neighbors in the direction $\alpha$ ($\alpha,\beta=x,y$) and
$\hat{n}_{\alpha{\bf j}}=\hat{a}_{\alpha{\bf j}}^\dagger\hat{a}_{\alpha{\bf j}}$ is the atom number
operator; and where the interaction terms  
\begin{equation}
H_{nn} = \sum_{\alpha}\sum_{\bf j} \frac{U_{\alpha
    \alpha}}{2}\hat{n}_{\alpha {\bf j}}\left(\hat{n}_{\alpha {\bf j}}-1\right) +\sum_{\alpha
  \beta, \alpha \neq \beta} \sum_{\bf j}  U_{\alpha
    \beta}\hat{n}_{\alpha{\bf j}}\hat{n}_{\beta{\bf j}}, 
\end{equation}
and
\begin{equation}\label{changefl}
\begin{array}{lll}
H_{FD} & = & \displaystyle{\sum_{\alpha \beta, \alpha \neq \beta} \sum_{\bf j}
\frac{U_{\alpha \beta}}{2} \left(\hat{a}^{\dagger}_{\alpha{\bf
    j}}\hat{a}^{\dagger}_{\alpha{\bf j}}\hat{a}_{\beta{\bf
    j}}\hat{a}_{\beta{\bf j}}\right.}\\ \\ 
& & +\left.\hat{a}^{\dagger}_{\beta{\bf j}}\hat{a}^{\dagger}_{\beta{\bf
    j}}\hat{a}_{\alpha {\bf j}}\hat{a}_{\alpha, {\bf j}}\right), 
\end{array}    
\end{equation}
account, respectively, for contribution of density-density and
interflavor conversion interactions. The expression for the
interaction parameters is given by
\begin{equation}\label{intpara}
U_{\alpha \beta} = U_0 \int d\vec{r}\, |w_{\alpha {\bf
    j}}(\vec{r})|^2|w_{\beta {\bf j}}(\vec{r})|^2,
\end{equation}
and for the tunneling coefficients by 
\begin{equation}\label{tunnelpara}
t_{\alpha\beta} = - \int d\vec{r}\,w_{\alpha {\bf
    j}}^*(\vec{r})\left[-\nabla^2 + V(\vec{r})\right]w_{\alpha
{\bf j + 1}_{\beta}}(\vec{r}),
\end{equation}
where by ${\bf j + 1}_{\beta}$ we indicate the neighboring site of
${\bf j}$ in the direction $\beta$, and $U_0=\tilde{U}_0l^3/E_r$, 
$V_\beta=\tilde{V}_\beta/E_r$ are the dimensionless
interparticle strength and lattice amplitudes, respectively. From
here, after substitution of the Wannier functions~(\ref{flavor_wfx})
into the above equation (\ref{tunnelpara}), it is straightforward to
see that contributions for the tunneling coefficient in the
direction perpendicular to the node depend uniquely from Wannier functions
of the first band (i.e. $\nu=1$), while in the direction of the node
it solely depends on the second band Wannier functions ($\nu=2$). As
a consequence, an $x$-flavor atom has larger probability of tunneling
in the $x$-direction than in the $y$-direction, while the opposite
also holds for a $y$-flavor atom. We continue discussions
regarding the effects of this anisotropic tunneling 
in Sec.~\ref{sec:sec2}. Also, before proceeding with the mean-field derivations, we make a
  brief comment on the symmetries of the
  Hamiltonian~(\ref{manyham}). As pointed out in Ref.~\cite{isaksson},
this Hamiltonian has an associated $Z_2$ symmetry, related to the
parity of atomic flavors: since atom scattering processes occur in
pairs, the number of $x$-flavor atoms $N_x$ and $y$-flavor atoms $N_y$
are preserved modulo 2. Isotropic lattices support, in addition,
a symmetry corresponding to swapping of atomic flavors
$x\leftrightarrow y$. We will also discuss how this property
implies a double degeneracy of the ground state for the infinite system.

\subsection{Mean-field Hamiltonian}
Except when otherwise stated, all our results follow from analysis of
the 2D lattice. We assume the condensate confined in the
transverse $z-$direction, and thus at each lattice site the system could 
either be purely two-dimensional or
form condensed tubes with typically a few hundred of
atoms~\cite{sengstock}. In either configuration, a mean-field treatment
is expected to give a reliable picture of the relevant physics~\cite{ph4}.

At a mean-field level, the operators $\hat{a}_{\alpha \vecj}$ are
replaced by the complex numbers $\psi_{\alpha \vecj}$. This
approximation is equivalent to assigning a coherent state at each
site, $|\Psi\rangle = \bigotimes_{\bf j}|\psi\rangle_{\bf j}=\bigotimes_{\bf j}|\psi_{x{\bf j}},\psi_{y{\bf
    j}}\rangle_{\vecj}$ such that $\hat{a}_{\alpha{\bf
    j}}|\Psi\rangle=\psi_{\alpha{\bf j}}|\Psi\rangle$. In terms of the
Fock basis, the single site many-body wavefunction reads 
\begin{equation}
|\psi\rangle_{\vecj} = \exp\left(-\frac{|\psi_{x\vecj}|^2 + |\psi_{y\vecj}|^2}{2}\right)\sum_{n_x, n_y}\frac{\psi_{x\vecj}^{n_x}\psi_{y\vecj}^{n_y}}{\sqrt{n_x!n_y!}}|{\bf n}\rangle_{\vecj},
\end{equation} 
where ${|\bf n\rangle}_{\vecj} = |n_x, n_y\rangle_{\vecj}$ represents
the state of $n_x$ $x$-flavor atoms and $n_y$ $y$-flavor atoms at site
${\bf j}$. Moreover, in this language the onsite order parameter of site ${\bf j}$ and flavor $\alpha$ reads $\psi_{\alpha{\bf  j}}=\langle\Psi|\hat{a}_{\alpha{\bf
    j}}|\Psi\rangle$.

With the coherent state ansatz we can obtain the equations of
motion for the order parameter $\psi_{\alpha{\bf j}}$ from the Euler-Lagrange
equations 
\begin{equation}
\frac{\partial L}{\partial \psi^{*}_{\alpha\vecj}}
-\frac{d}{dt}\left(\frac{\partial L}{\partial\dot{\psi^{*}_{\alpha\vecj}}}\right) = 0,
\end{equation}
where the Lagrangian is given by
\begin{equation}
L = \sum_\alpha\sum_{\vecj} i\frac{1}{2}\left[\psi^{*}_{\alpha\vecj}\frac{d}{dt}\psi_{\alpha\vecj} - \psi_{\alpha\vecj}\frac{d}{dt}\psi^{*}_{\alpha\veci}\right] - H_{MF},
\end{equation}
with the mean-field Hamiltonian 
\begin{equation}
\begin{array}{lll}
H_{MF} & = & \displaystyle{-\sum_{\alpha, \beta}\sum_{\langle\veci\vecj\rangle_{\alpha}}t_{\alpha
\beta}\psi^{*}_{\alpha\veci}\psi_{\alpha\vecj} +
\sum_{\alpha}\sum_{\vecj}\frac{U_{\alpha\alpha}}{2}n_{\alpha\vecj}n_{\alpha\vecj}}\\ \\
& & +\displaystyle{\sum_\alpha\sum_{\bf j}\frac{\omega^2}{2}\left(x_{\bf j}^2+y_{\bf j}^2\right)n_{\alpha{\bf j}}}\\ \\
& & \displaystyle{+\sum_{\alpha\beta,
  \alpha\neq\beta}\sum_{\vecj}U_{\alpha\beta}n_{\alpha\vecj}n_{\beta\vecj} +
\sum_{\alpha\beta,\alpha\neq\beta}\sum_{\bf j}\frac{U_{\alpha\beta}}{2}}\\ \\
& & \times\left(\psi^{*}_{\alpha\vecj}\psi^{*}_{\alpha\vecj}\psi_{\beta\vecj}\psi_{\beta\vecj} + \psi^{*}_{\beta\vecj}\psi^{*}_{\beta\vecj}\psi_{\alpha\vecj}\psi_{\alpha,\vecj}\right),
\end{array}
\end{equation}
and where the  Hamiltonian~(\ref{manyham}) has been
normally ordered prior to calculation of the coherent state expectation
value. Here the density of the flavor $\alpha$ is given by $n_{\alpha{\bf j}} =
\vert\psi_{\alpha\vecj}\vert^2$ and normalization was imposed in the
whole lattice as
\begin{equation}
N=N_x+N_y=\sum_{\bf j}|\psi_{x{\bf j}}|^2+\sum_{\bf j}|\psi_{y{\bf j}}|^2,
\end{equation}
with $N$ accounting for the total number of atoms.

The Euler-Lagrange equations then correspond to a set of coupled Gross-Pitaevskii equations,
one for each atomic $\alpha$-flavor at each site ${\bf j}$: 
\begin{equation}\label{disc-gp}
\begin{array}{lll}
\displaystyle{i\frac{\partial \psi_{x\vecj}}{\partial t}} & = & \displaystyle{-\sum_{\beta \in
  \{x, y\}}t_{x\beta}\left(\psi_{x{\bf j + 1}_{\beta}} - 2\psi_{x\veci} + \psi_{x{\bf j - 1}_{\beta}}\right)} \\ \\
& & +\displaystyle{\frac{\omega^2}{2}\left(x_{\bf j}^2+y_{\bf j}^2\right)\psi_{x{\bf j}}}\\ \\
& & +\left(U_{xx}|\psi_{x\vecj}|^2 + 2U_{xy}|\psi_{y\vecj}|^2\right)\psi_{x\vecj}\\ \\
& & + (U_{xy} + U_{yx})\psi_{y\vecj}^2\psi_{x\vecj}^*\\ \\
\displaystyle{i\frac{\partial \psi_{y\vecj}}{\partial t}} & = & \displaystyle{-\sum_{\beta \in
  \{x, y\}}t_{y\beta}\left(\psi_{y{\bf j + 1}_{\beta}} - 2\psi_{y\veci} + \psi_{y{\bf j - 1}_{\beta}}\right)} \\ \\
& & +\displaystyle{\frac{\omega^2}{2}\left(x_{\bf j}^2+y_{\bf j}^2\right)\psi_{y{\bf j}}}\\ \\
& & +\left(U_{yy}|\psi_{y\vecj}|^2 + 2U_{yx}|\psi_{x\vecj}|^2\right)\psi_{y\vecj}\\ \\
& & + (U_{yx} + U_{xy})\psi_{x\vecj}^2\psi_{y\vecj}^*.
\end{array}
\end{equation}
Like all other parameters and variables, time $t$ is a dimensionless
quantity. Note also that we take all the
  parameters entering the above equations from numerically obtained
  Wannier overlap integrals according to Eqs.~(\ref{intpara}) and
  (\ref{tunnelpara}), and consequently no harmonic approximation is
  imposed. This avoids some qualitatively wrong conclusions
  which can occur with the latter assumption~\cite{pinf2}. 

\section{Ideal gas}\label{sec:sec2}

\subsection{Ground state properties}
Let us first investigate some features of the
system in the non-interacting case, where the free mean-field
Hamiltonian  is given by
\begin{equation}\label{freeham}
\begin{array}{lll}
H_{MF}^{(0)} & = &\displaystyle{-\sum_{\alpha, \beta}\sum_{\langle\veci\vecj\rangle_{\alpha}}t_{\alpha
\beta}\psi^{*}_{\alpha\veci}\psi_{\alpha\vecj}}\\ \\
& & \displaystyle{ +\sum_\alpha\sum_{\bf j}\frac{\omega^2}{2}\left(x_{\bf j}^2+y_{\bf j}^2\right)n_{\alpha{\bf j}}.}
\end{array}
\end{equation}
In the absence of interflavor interactions, interflavor onsite
coherence is not established. However, within each flavor it is the
tunneling which determines how the phases of neighboring sites are
related to each other. We thus characterize these properties for the ground-state by minimizing
(\ref{freeham}). To this end, the expression for the onsite order
parameters is taken as $\psi_{\alpha{\bf j}}=|\psi_{\alpha\vecj}|e^{i\phi_{\alpha\vecj}}$,
and by noticing that $t_{xx},\; t_{yy} < 0$ and $t_{xy},\;t_{yx} > 0$
we obtain a striped order in the phase of each flavor. More explicitly,
the phase of the $x$-flavor order parameter can be expressed as
$\phi_{x{\bf j}}=\phi_{x}(j_x,j_y)=\pi\times\!\!\mod(j_x,2)$.
This means that neighboring sites will
always keep the same phase in the
direction perpendicular of the node, while in the parallel direction
the phase difference will be $\pi$.

The discrete model (\ref{freeham}) can in principle be solved analytically by noticing that the Hamiltonian matrix has the same structure as the one of the Mathieu equation expanded in momentum eigenstates~\cite{mathieu}. The solutions is not very instructive as it is determined from the Fourier expansion of the Mathieu functions, i.e. by the transformation matrix between quasi- and real momentum. A simple physical picture of the influence of the trap in the discrete model is instead better analyzed in the continuum limit where the analytical solutions can be given in closed forms. Here it is convenient to work with the order parameters without phase modulation. We thus impose the correct 
phase imprint responsible for rendering the striped order into the
wavefunction ansatz. Under these circumstances, the phase factors
can be absorbed into the redefinition of the tunneling coefficient,
$t_{\alpha\alpha}\rightarrow-t_{\alpha\alpha}$. 
In addition, the continuum limit consists 
in $\psi_{\alpha{\bf j}}\rightarrow\psi_\alpha(x,y)$, 
and the kinetic energy transforms
as 
\begin{equation}
\psi_{\alpha{\bf j + 1}_{\beta}} - 2\psi_{\alpha\veci} + \psi_{\alpha{\bf j - 1}_{\beta}}\longrightarrow\frac{\partial^2}{\partial\beta^2}\psi_\alpha(\alpha,\beta).
\end{equation}
With this approximation, we obtain the following continuum equations 
\begin{equation}\label{non_int_1}
\begin{array}{lll}
\displaystyle{i\frac{\partial}{\partial t}\psi_{x}(x,y)} & = &  \displaystyle{
\left[-|t_{xx}|\frac{\partial^2}{\partial x^2}-|t_{xy}|\frac{\partial^2}{\partial y^2}\right.}\\ \\
& & \displaystyle{ \left.+\frac{\omega^2}{2}\left(x^2+y^2\right)\right]\psi_{x}(x,y)},\\ \\
\displaystyle{i\frac{\partial}{\partial t}\psi_{y}(x,y)} & = & \displaystyle{\left[-|t_{yy}|\frac{\partial^2}{\partial y^2}-|t_{yx}|\frac{\partial^2}{\partial x^2}\right.}\\ \\
& & \displaystyle{ + \left.\frac{\omega^2}{2}\left(x^2+y^2\right)\right]\psi_{y}(x,y)},
\end{array}
\end{equation}
where $x$ and $y$ are dimensionless. By introducing the effective
mass $m_{\alpha\beta}=|t_{\alpha\beta}|^{-1}/2$ and parallel and
transverse frequencies 
\begin{equation}\label{efffreq}
\begin{array}{llll}
\displaystyle{\omega_\parallel=\omega\sqrt{2|t_{\alpha\beta}|}}, & & & \alpha\neq\beta,\\ \\
\displaystyle{\omega_\perp=\omega\sqrt{2|t_{\alpha\beta}|}}, & & & \alpha=\beta,
\end{array}
\end{equation}
Eq.~(\ref {non_int_1}) can be written as 
\begin{equation}\label{contin2}
\begin{array}{lll}
\displaystyle{i\frac{\partial}{\partial t}\psi_x(x,y)} & = & \displaystyle{\left[\frac{p_x^2}{2m_{xx}}+\frac{p_y^2}{2m_{xy}}\right.}\\ \\
& & \displaystyle{\left.+\frac{m_{xx}\omega_{\parallel}^2}{2}x^2+\frac{m_{xy}\omega_{\perp}^2}{2}y^2\right]\psi_\alpha(x,y),}
\end{array}
\end{equation} 
with a similar equation for the $y$-flavor. We find,
therefore, that the continuum approximation reduces the
system to two 2D anisotropic harmonic
oscillators. It is important to stress though, that in
order to derive Eq.~(\ref{contin2}), the striped order 
must be correctly implemented.  If the phase modulation is not
considered before imposition of the continuum approximation,
the resulting Hamiltonian is not bounded from below, and since the
lattice naturally introduces a momentum 
cut-off $\Lambda=\pi/\lambda$ at the edges of
the Brillouin zone, it is a property not present in the discrete
model. The initial phase imprint is thus a tool to circumvent
this problem, where the overall effect of the procedure translates into
inversion of the $p$-band and shifting of its minimum to the center of
the Brillouin zone.

In the continuum model, the anisotropy arising from the different
tunneling elements $t_{xx}$ and $t_{xy}$ is directly reflected in the
direction-dependence of $m_{\alpha\beta}$ and
$\omega_{\alpha\beta}$. Therefore, it follows from this anisotropy that the
continuum Gaussian ground state will have different widths in the two
directions $x$ and $y$. We use this fact to define the anisotropy parameter
\begin{equation}\label{squeez}
S_{x} =\sqrt{\frac{(\Delta_xx)^2}{(\Delta_xy)^2}},
\end{equation}
with equivalent expression to the $y$-flavor, and where
$(\Delta_\alpha\beta)^2=\langle\beta^2\rangle_\alpha-\langle\beta\rangle_\alpha^2$ 
and $\langle\cdots\rangle_\alpha$ represents the expectation value
taken with respect to $\psi_\alpha(x,y)$.  
For symmetry reasons $S_xS_y=1$ must hold, and thus we call the
$x$-flavor anisotropy parameter simply by $S$. 
This definition is general and applies to both the discrete as
well as for the continuum limit. It can be used to derive an explicit expression
for the latter case
\begin{equation}\label{sqcon}
S_{con}=\left(\frac{|t_{xx}|}{|t_{xy}|}\right)^{1/4}=\left(\frac{\omega_\parallel}{\omega_\perp}\right)^{1/2}, 
\end{equation}
which as expected, predicts $S=1$ for isotropic systems (i.e., where
both directions have the same tunneling strengths). However, 
generally $S \neq 1$, and therefore it reveals the existence of narrowing
in the flavor density along one of the directions. This anisotropy is
a consequence of the direction-dependence of the tunneling
$t_{\alpha\beta}$ and is a result beyond the LDA. Note furthermore that when atom-atom interaction has
been neglected, $N_x$ and $N_y$ are preserved quantities and the
actual ground state of the system will be determined from the
preparation process. For non-zero atom-atom interaction, $N_x$ and
$N_y$ are no longer independently preserved due to the term
(\ref{changefl}) and the interaction energy is minimized with
$N_x=N_y$ as will be seen in the next section. Now we continue with
further discussions upon validity and applicability of the continuum
approximation.  

\subsection{Ideal gas at finite temperatures}
For the ideal gas system, represented by the Hamiltonian
(\ref{freeham}), it is rather straightforward to calculate finite
temperature effects, either from direct numerical diagonalization or using the analytical solutions obtained from Fourier expansions of Mathieu functions. Due to discretization of (\ref{freeham}), the
eigenstates in the harmonic trap are not the same as the  
usual eigenstates of the harmonic oscillator. Since implications of
this for the thermodynamics of an ideal gas are not clear, we
numerically solve the discrete 2D and also 3D Schr\"{o}dinger
equations for the eigenstates, and use these as a basis to
study Bose-Einstein condensation on the $p$-band in the presence of a trap.

In the continuum limit described by Eq.~(\ref{non_int_1}), the critical temperature for
the Bose-Einstein condensation in the harmonic trap is well
known~\cite{Pethick2001a} and given by 
\beq
T_{c0}^{(2D)}=\omega_{eff}^{(2D)}\sqrt{6N/\pi^2}
\enq
in 2D and in 3D by
\beq
T_{c0}^{(3D)}=\omega_{eff}^{(3D)}\left(N/\zeta(3)\right)^{1/3},
\enq
with $\zeta(3)\approx 1.20206$, and where the trapping frequencies are
defined as averages of the effective frequencies~(\ref{efffreq}) as  
\beq
\omega_{eff}^{(2D)}=4\omega\sqrt{|t_{xx}| |t_{xy}|} 
\enq
and
\beq
\omega_{eff}^{(3D)}=4\omega\left(|t_{xx}| |t_{xy}|^2\right)^{1/3}.
\enq
For bosonic gases, the number $N_{T}$ of thermal
(non-condensed) atoms follows from 
\beq
N_{T}=\sum_{n\neq 0} \frac{1}{\exp(\beta \left(E_n-\mu\right))-1},
\enq
where $\beta=E_r/k_BT$ is the inverse (dimensionless) temperature and $\mu$ is the
chemical potential. Together with the eigenenergies  $E_n$ obtained by solving the
Schr\"{o}dinger equation, this can be used to compute
the critical temperature for condensation in our lattice model. Notice
however, that while below the critical temperature the chemical potential $\mu$
is equal to the ground state energy, at higher temperatures it must be
determined by fixing the total atom number to $N$. 

\begin{figure}
\includegraphics[width=0.45\textwidth]{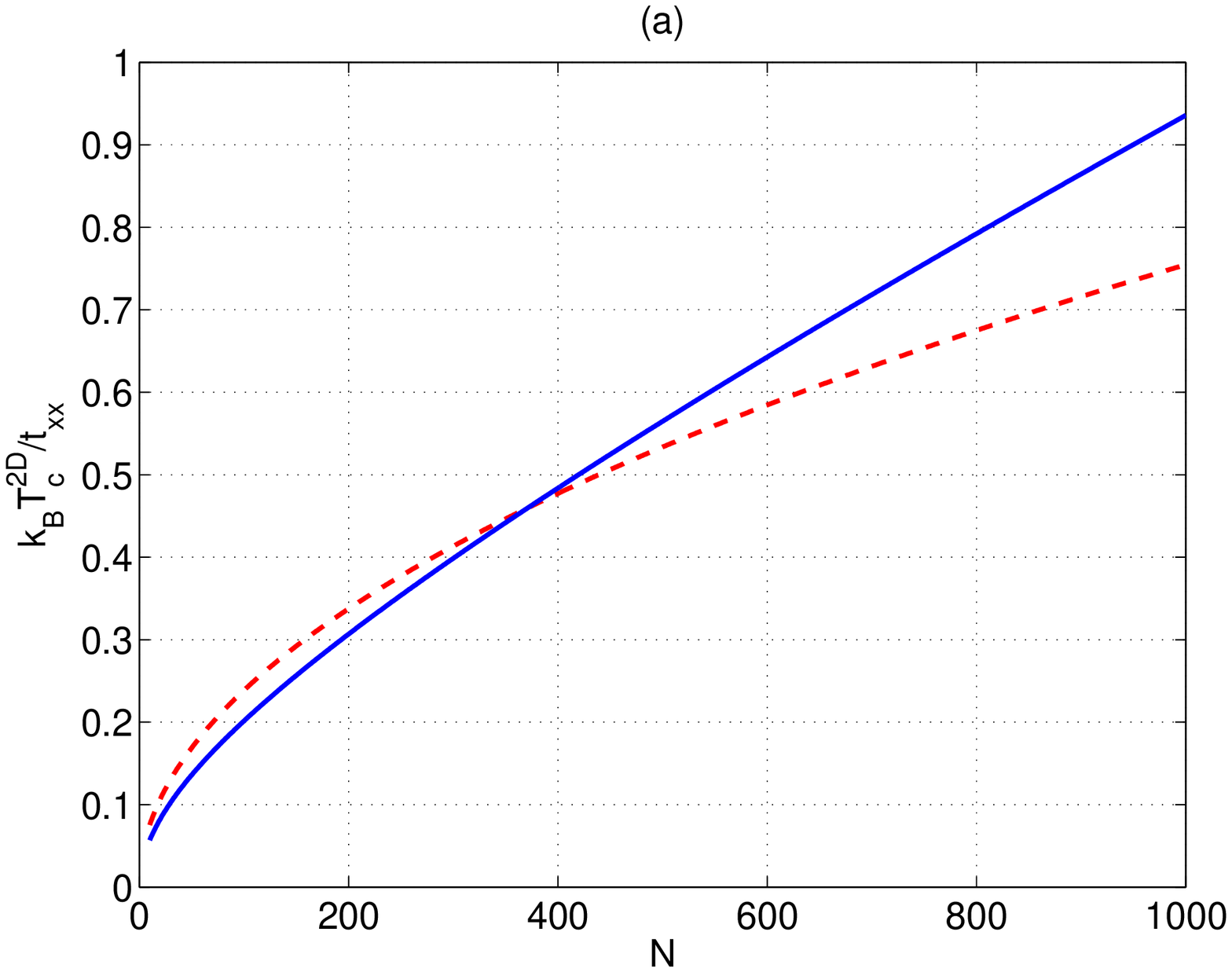} 
\includegraphics[width=0.45\textwidth]{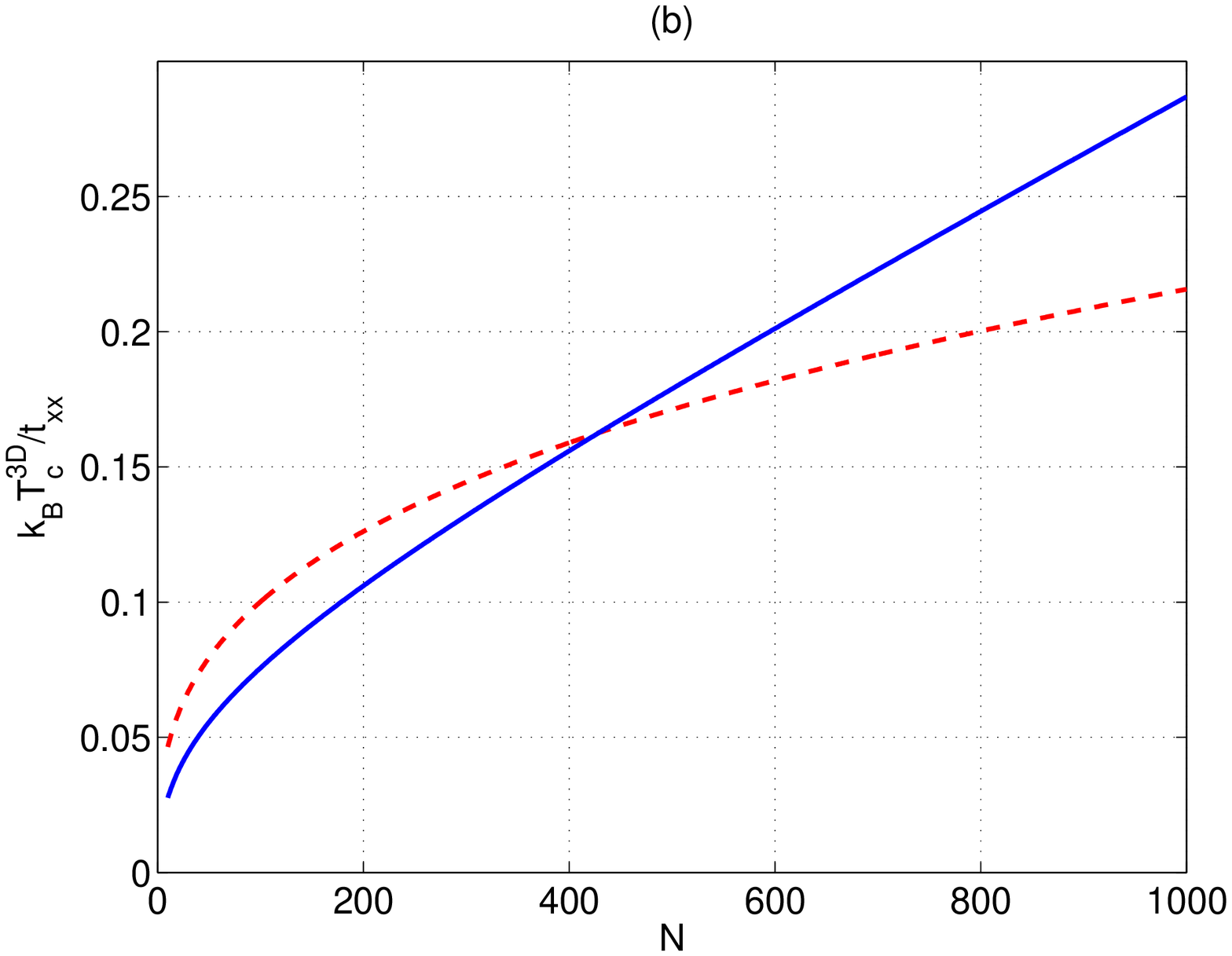}
\caption{(Color online) The critical temperature for the Bose-Einstein 
condensation as a function of atom number $N$ in
(a) 2D system and (b) 3D system.
The dashed line shows the result based on approximating the
discrete model with a continuum one, and the solid line displays the numerically calculated results of the discrete model. We used the dimensionless trap
strength $\omega^2/2=0.001$ and $|t_{xx}/t_{xy}|=20.1$  which is our
estimate for the ratio of tunneling strengths at $V_x=V_y=17$.}
\label{fig:IdealTc}
\end{figure}

We compare the predictions for the critical temperature of the continuum
and lattice models in
Fig.~\ref{fig:IdealTc}. As is seen, the general result in both the 2D
and 3D systems, consists in a somewhat lower
critical temperature for very small atom numbers, but substantially
larger critical temperature for high atom numbers. Such difference is
due to different density of states between the lattice 
and the continuum models. 

\begin{figure}
\includegraphics[width=0.45\textwidth]{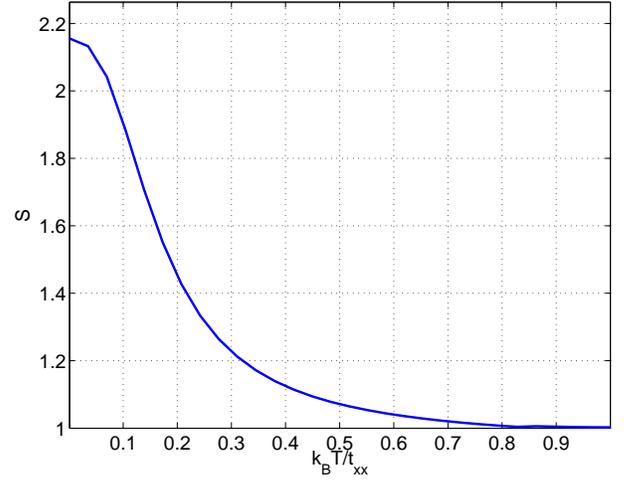} 
\caption{(Color online) The anisotropy parameter $S$ of the 2D density
  distribution as a function of the $t_{xx}$-scaled temperature for $1000$ atoms,
  dimensionless trap strength $\omega^2/2=0.001$  and potential depth $V_x=V_y=17$.}
\label{fig:aniso}
\end{figure} 

\begin{figure}
\includegraphics[width=0.45\textwidth]{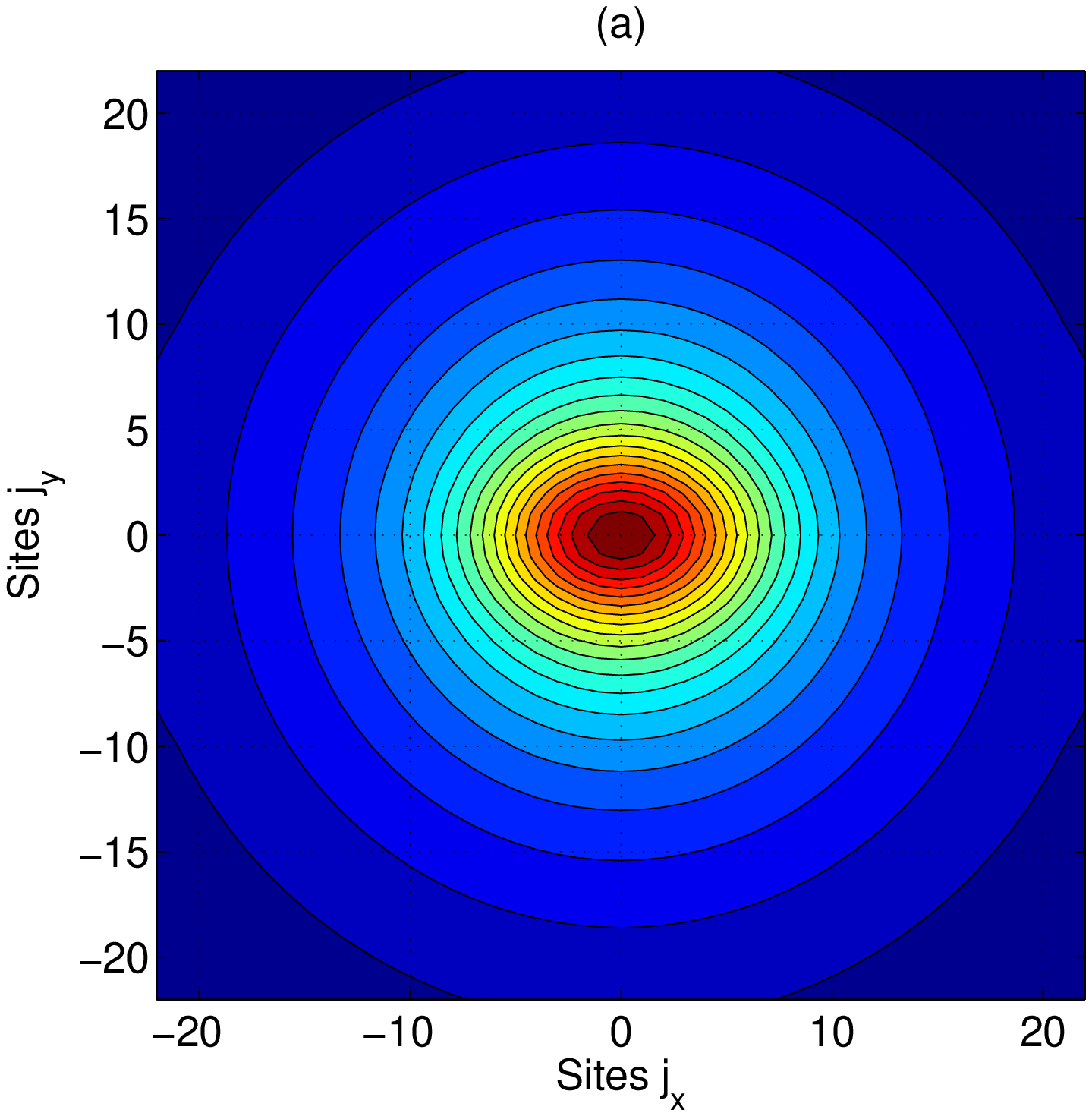} 
\includegraphics[width=0.45\textwidth]{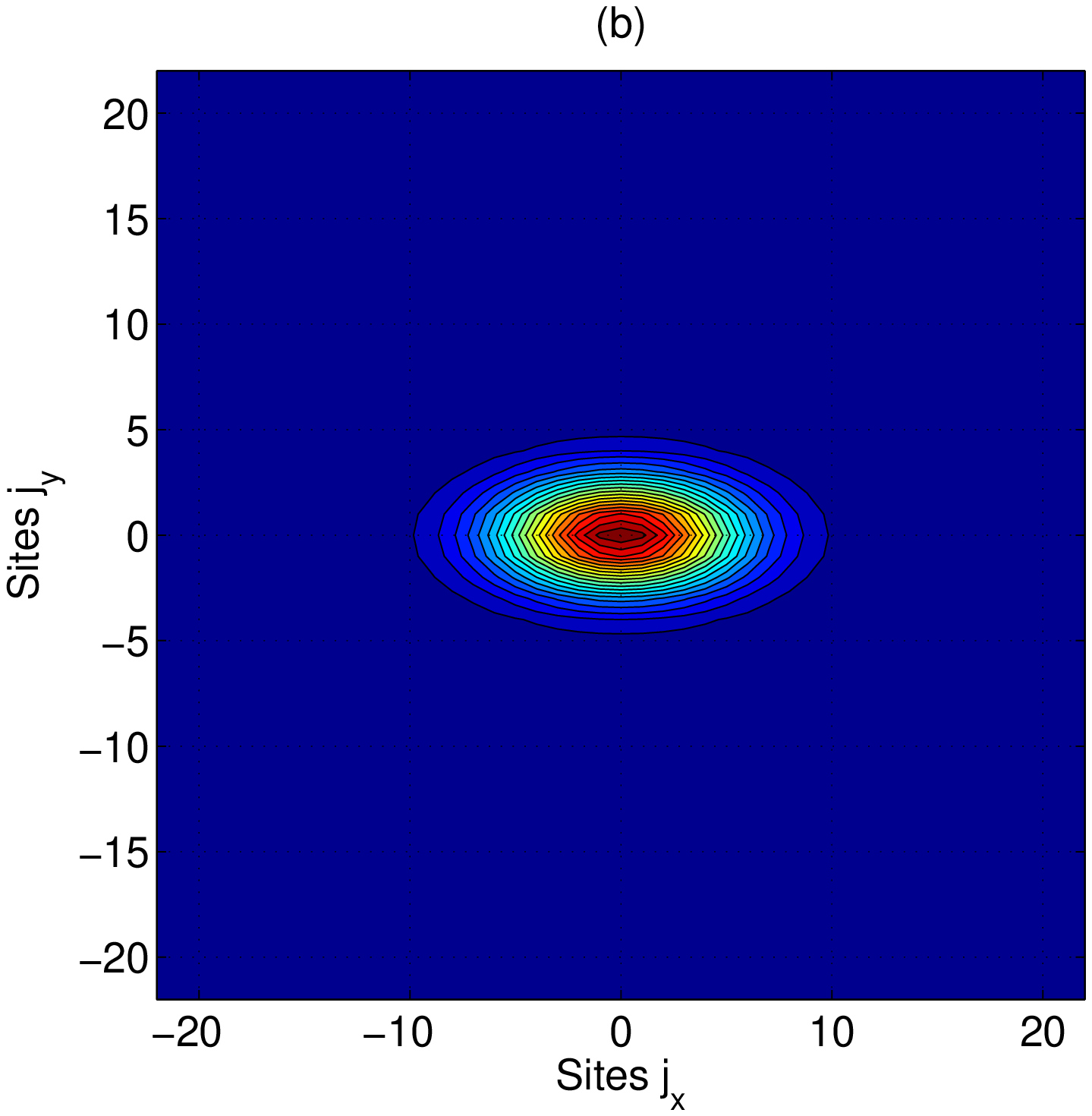}
\caption{(Color online) The populations per site (for a single atomic flavor) of the 2D
Bose gas close ($k_BT/t_{xx}=1$) to the condensation critical
temperature (a), and at $T=0$ (b). In both examples, the number of
atoms is $1000$, dimensionless trapping strength $\omega^2/2=0.001$, 
and potential depth $V_x=V_y=17$.}
\label{fig:densT}
\end{figure}

In a trap, the transition to the condensed state is
typically associated with pronounced changes in the atomic density
distribution. A broad thermal distribution above the critical
temperature acquires a bimodal structure as a density peak appears
in the center corresponding to the macroscopic occupation
of the condensate ground state. 
Also, as already discussed in the previous subsection, in the case of trapped $p$-band
atoms the anisotropy is a new feature appearing in the
density distribution. Above the critical temperature $T_c$, the density distribution has 
the same width in $x$- and $y$-directions, but below $T_c$ the
condensate density distribution shares the properties
of the ground state, which
is anisotropic due to different tunneling strengths in different directions.
We give an example of this behavior in Fig.~\ref{fig:aniso} by
displaying the anisotropy parameter~(\ref{squeez}) as a function of
temperature for $1000$ atoms. Furthermore, 
in Fig.~\ref{fig:densT} we show the density  ($\psi_n({\bf j})$
are the eigenstate wavefunctions)
\beq
n_{tot}({\bf j})=N_0|\psi_0({\bf j})|^2+\sum_{n\neq 0} 
\frac{|\psi_n({\bf j})|^2}{\exp(\beta \left(E_n-\mu\right))-1}
\enq
close to $T_c$ and at $T=0$, demonstrating 
the appearance of strong anisotropy (for single flavor) at low temperatures.

\section{Interacting gas}\label{sec:sec4}
\subsection{Characterizing the ground state}
Until now we have not considered how interactions affect
the system's ground state properties. Effects stemming from the
tunneling part and the corresponding phase
ordering imposed in the minimization of the mean-field Hamiltonian
were already discussed in Sec.~\ref{sec:sec2}. We thus complete the
characterization of the ground state of the system by repeating this
analysis to the interacting part of $H_{MF}$. Since neighboring sites
are not coupled by the interaction term, it is  enough to consider the
energy contribution within only one single site. In analogous
procedure to the one used in the aforementioned analysis, we substitute
the expression $\psi_{\alpha{\bf j}}=|\psi_{\alpha{\bf
    j}}|e^{i\phi_{\alpha{\bf j}}}$ for the onsite order parameter of
the flavor $\alpha$, and the resulting density-density and interflavor 
conversion parts of the mean-field Hamiltonian follow, respectively, as
\begin{equation}
H_{nn}^{({\bf j})} = \frac{U_{xx}}{2}|\psi_{x\vecj}|^4 +
\frac{U_{yy}}{2}|\psi_{y\vecj}|^4 + (U_{xy} + U_{yx})|\psi_{x\vecj}|^2|\psi_{y\vecj}|^2
\end{equation}
and
\begin{equation}
H_{FD}^{({\bf j})} =
\frac{U_{xy} + U_{yx}}{2}|\psi_{x\vecj}|^2|\psi_{y\vecj}|^2\cos(2(\phi_{x\vecj} -
\phi_{y\vecj})). 
\end{equation}
Here, the term accounting for the density-density interactions is phase
independent and gives no information about the on-site
phase ordering. However, the interflavor conversion term will
explicitly depend on the phase difference between the $x$- and
$y$-flavor order parameters, and accordingly, establishes an onsite interflavor 
phase locking. In fact, when $U_{xy}, U_{yx} > 0$
the onsite energy is minimized with
$\phi_{x\vecj}-\phi_{y\vecj} = \pm\pi/2$.

Now combining the above argument with the results of
Sec.~\ref{sec:sec2}, we obtain both the on- and inter-site full phase
coherence of the condensate within the lattice. To this end we adopt
the position representation of the onsite order parameter 
\begin{equation}\label{onsiteorder2}
\psi_{\bf j}(\vec{r}) = \psi_{x{\bf j}}w_{x{\bf j}}(\vec{r}) + \psi_{y{\bf j}}w_{y{\bf j}}(\vec{r})
\end{equation}
and apply the requirements of phase locking, which yield
\begin{equation}\label{onsiteorder}
\psi_{\bf j}(\vec{r}) = |\psi_{x{\bf j}}|w_{x{\bf j}}(\vec{r}) \pm i |\psi_{y{\bf j}}|w_{y{\bf j}}(\vec{r}),
\end{equation}
where the $\pm$-sign alternates between neighboring sites. 
Note that in the absence of a trap, flipping the sign on all the sites
gives a new configuration with exactly the same energy. This
characteristic degeneracy, related to the swapping of the flavors
$x\leftrightarrow y$, was already pointed out
earlier. By furthermore considering the orthonormality property of
Wannier functions, $\int\,d\vec{r}w_{\alpha{\bf
    j}}^*(\vec{r})w_{\beta{\bf
    i}}(\vec{r})=\delta_{\alpha\beta}\delta_{{\bf ji}}$, we interpret the onsite order parameter as a spinor  
\begin{equation}\label{spinorder}
\psi_{\bf j}=\left[
\begin{array}{c}
|\psi_{x{\bf j}}|\\
\pm i|\psi_{y{\bf j}}|
\end{array}\right],
\end{equation}
where the spatial dependence has been absorbed into the basis states
$w_{x{\bf j}}(\vec{r})$ and $w_{y{\bf j}}(\vec{r})$. In particular,
the length of the spinor defined in this way gives the number of atoms
at site ${\bf j}$, i.e. $N_{\bf   j}=\sqrt{|\psi_{x{\bf j}}|^2+|\psi_{y{\bf
      j}}|^2}$. For having the same properties as a two-level system,
 the spinor onsite order parameter can be fully characterized by
the Bloch vector ${\bf J}_{\bf j}=(J_{x{\bf j}},J_{y{\bf j}},J_{z{\bf
    j}})$, where the components are 
\begin{equation}
\begin{array}{l}
J_{x{\bf j}} = \psi_{x{\bf j}}^{*}\psi_{y{\bf j}} + \psi_{y{\bf j}}^{*}\psi_{x{\bf j}},\\ \\
J_{y{\bf j}} = i\left(\psi_{x{\bf j}}^{*}\psi_{y{\bf j}} - \psi_{y{\bf j}}^{*}\psi_{x{\bf j}}\right),\\ \\
J_{z{\bf j}} = |\psi_{x{\bf j}}|^2 - |\psi_{y{\bf j}}|^2.
\end{array}
\end{equation}
In this picture, the length of the Bloch vector $|{\bf J}_{\bf
  j}|=N_{\bf j}$ corresponds to the number of atoms at site ${\bf j}$,
$J_{z{\bf j}}$ is the population imbalance between the two
flavors, and due to the specific phase locking in (\ref{spinorder}), we have
$J_{x{\bf j}}=0$. We also point out that the Bloch vector
constructed here corresponds to a mean-field version of the Schwinger
angular momentum representation~\cite{schwing}. 

While the Bloch vector contains all the information about the spinor
order parameter (\ref{spinorder}), it does not contain the full
information on the spatial dependence of the onsite order parameter
(\ref{onsiteorder}). This can be most easily investigated in the
harmonic approximation, where the Wannier functions are replaced by
harmonic eigenstates. Using this description, we have 
\begin{equation}\label{HAorder}
\psi_{\bf 0}^{(ha)}(\vec{r}) = \left[|\psi_{x{\bf 0}}|x \pm i|\psi_{y{\bf 0}}|y\right]e^{-\frac{x^2+y^2}{\sigma}}
\end{equation}
with $\sigma$ being the effective width determined from the lattice
amplitude. It is clear that for $|\psi_{x{\bf j}}|=|\psi_{y{\bf j}}|$ the
above onsite order parameter represents a vortex/anti-vortex state
with quantization $L_{z{\bf j}}\psi_{\bf
  j}^{(ha)}(\vec{r})=\pm\psi_{\bf j}^{(ha)}(\vec{r})$ where $L_{z{\bf
    j}}=-i\partial_{\phi_{\bf j}}$. This is only true, however, in the
harmonic approximation and when $J_{z{\bf j}}=0$. Beyond the harmonic 
approximation this is not strictly true even when $J_{z{\bf
    j}}=0$. Nevertheless, due to the properties of the Wannier
functions, Eq.~(\ref{flavor_wfx}), a $\pi/2$ phase difference between flavors
implies that the
condensate density vanishes at the center of site ${\bf j}$ and that
the condensate has a vortex like singularity in it.

\subsection{Properties in the symmetric lattice}
In the previous subsection we introduced the quantities characterizing
the physical state within each site. For the global properties we use
the anisotropy parameter as defined in Eq.~(\ref{squeez}). We
numerically solve Eq.~(\ref{disc-gp}) by employing the
split-operator method~\cite{split}, which is based on
factorization of the time-evolution operator into spatial and
momentum parts. This implies that the method is exact only in 
the limit of vanishingly
small time step. Therefore, propagation is divided into
small time steps and we verify the numerical accuracy by varying their
size. In order to find the ground state we propagate an initial
trial state in imaginary time until convergence has been
reached. It is generally seen that convergence is faster if we assume an initial
guess with the phase ordering properties discussed in the previous
section. It is also important to notice that a poor choice for
the initial state may result in convergence to local, 
but not global, energy minimum. To avoid this, we compare many
different simulations were the initial trial state has been varied and
the one with lowest final energy is assumed to be the ground state. The size of the grid is taken such that the atomic population is
approximately zero at the edge of the grid, and in all simulations we
consider a 2D system. The parameters of the Hamiltonian are calculated using the
numerically obtained Wannier functions, and consequently we do not
impose the harmonic approximation. 

We have seen that the tunneling and the onsite interaction establish a
phase locking according to Eq.~(\ref{onsiteorder}). In a system
without the external trap and $U_0\neq0$, it follows that $J_{y{\bf
    j}}/N_{\bf j}$ will either be $+1$ or $-1$, and the system
possesses a checkerboard structure, i.e. an anti-ferromagnet state
with spins alternating between pointing in the positive or negative
$y$-direction. The condensate will thus show the staggered
vortex/anti-vortex structure. Within the validity of the tight-binding
and single-band approximations, this result is exact. However, the strict vortex quantization
$L_{z{\bf j}}\psi_{\bf j}^{(ha)}(\vec{r})=\pm\psi_{\bf
  j}^{(ha)}(\vec{r})$ is only precise in the harmonic
approximation. In the presence of the trap, the inhomogeneities in the
density together with the tunneling anisotropy typically give rise to
onsite interflavor population imbalance, which tends to break the
anti-ferromagnetic order and lower the onsite angular momentum per
particle from $1$, which is expected from a quantized vortex with angular
momentum along $z$.  

\begin{figure}
\includegraphics[width=0.4\textwidth]{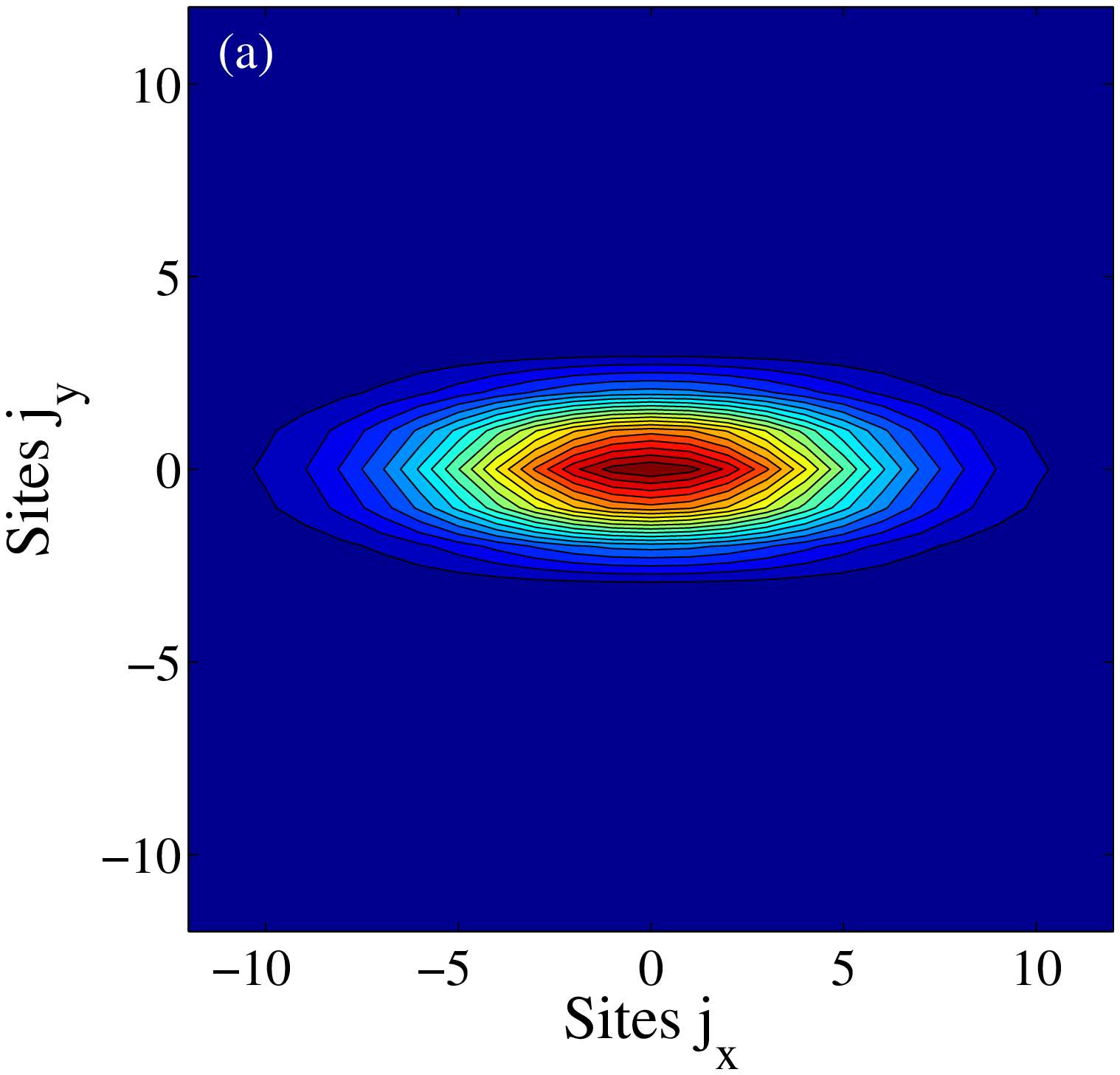}
\includegraphics[width=0.4\textwidth]{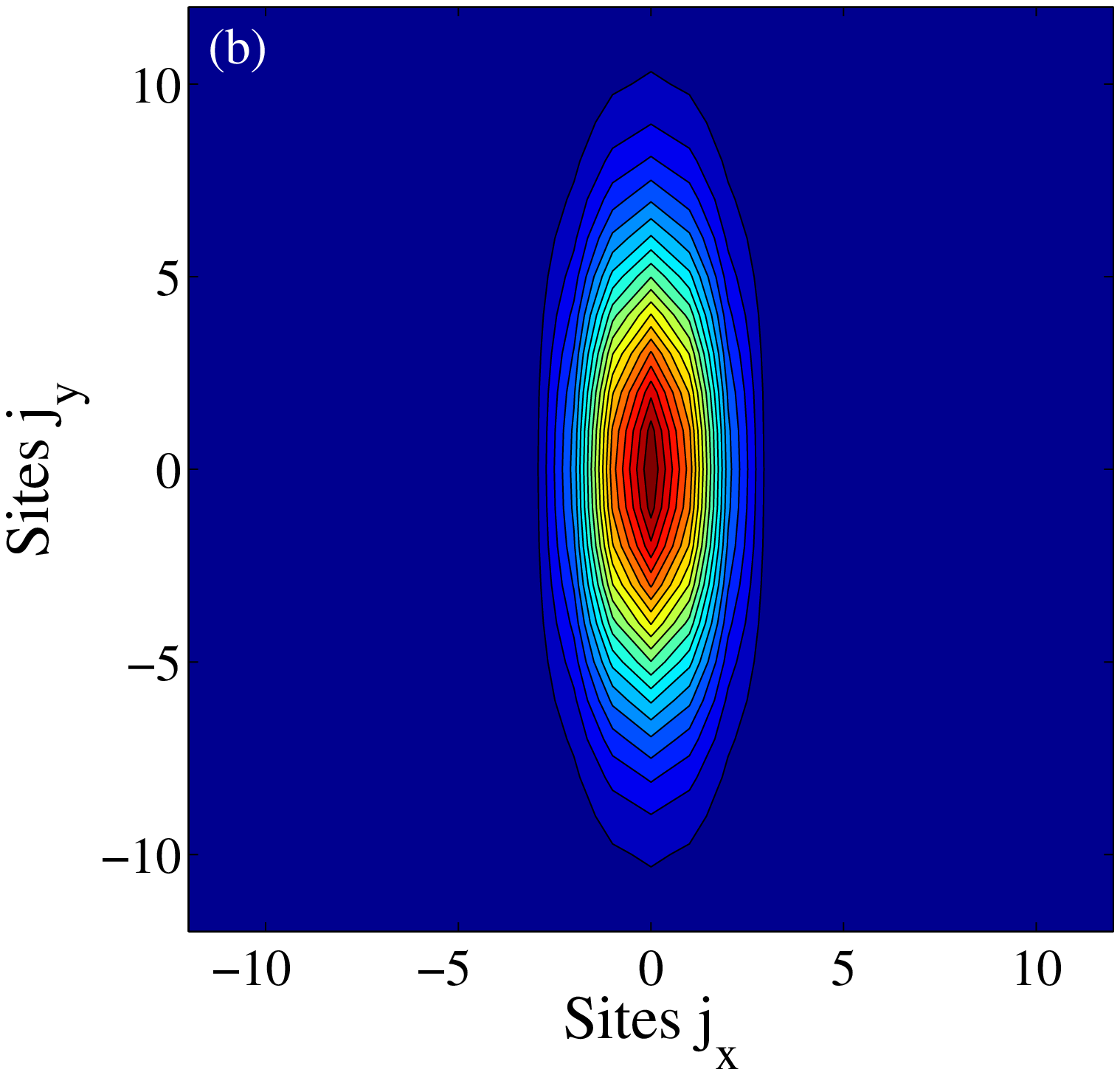}
\includegraphics[width=0.4\textwidth]{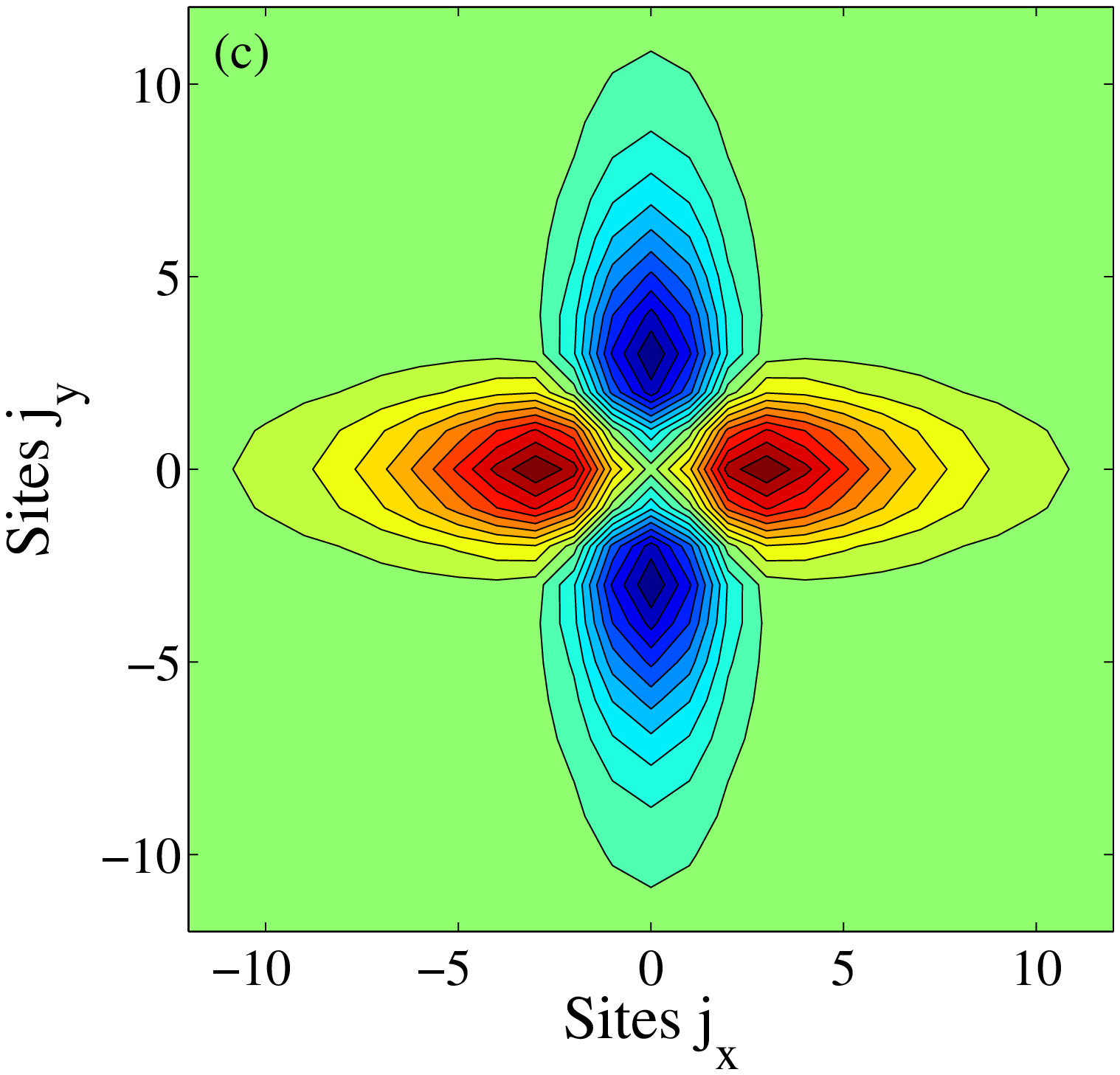}
\caption{(Color online) Upper two plots (a) and (b) show the $x$- and
  $y$-flavor ground state population respectively. The lower plot (c)
  gives instead the corresponding population imbalance $J_{z{\bf
      j}}$. The dimensionless system parameters are $V_x=V_y=17$,
  $\omega=0.005$, and $U_0N=1$. (Red color indicates an excess of
$x$-flavor atoms while blue regions have an excess of
$y$-flavor atoms.)} 
\label{latticepop}
\end{figure}

\begin{figure}
\includegraphics[width=0.45\textwidth]{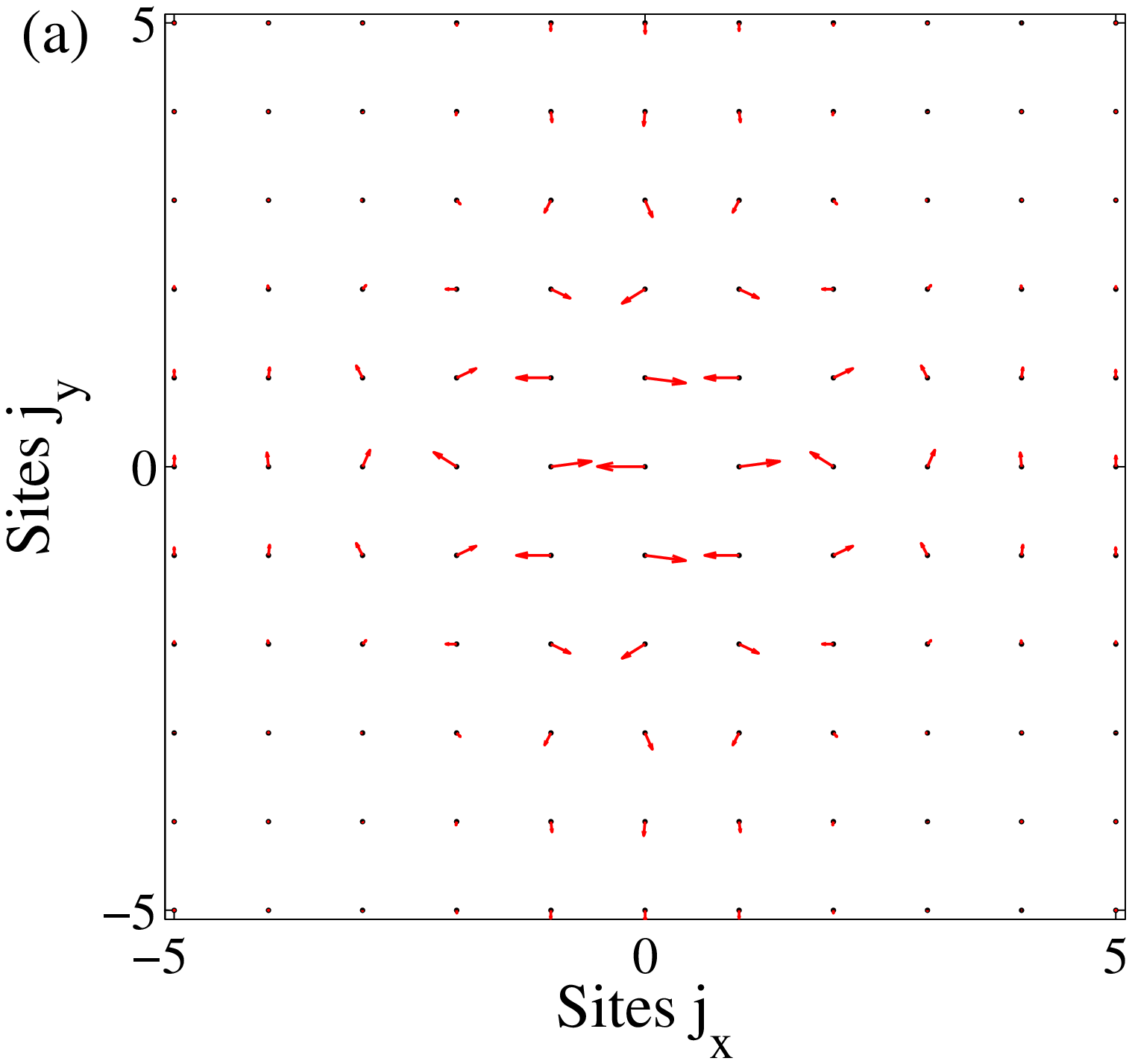}
\includegraphics[width=0.45\textwidth]{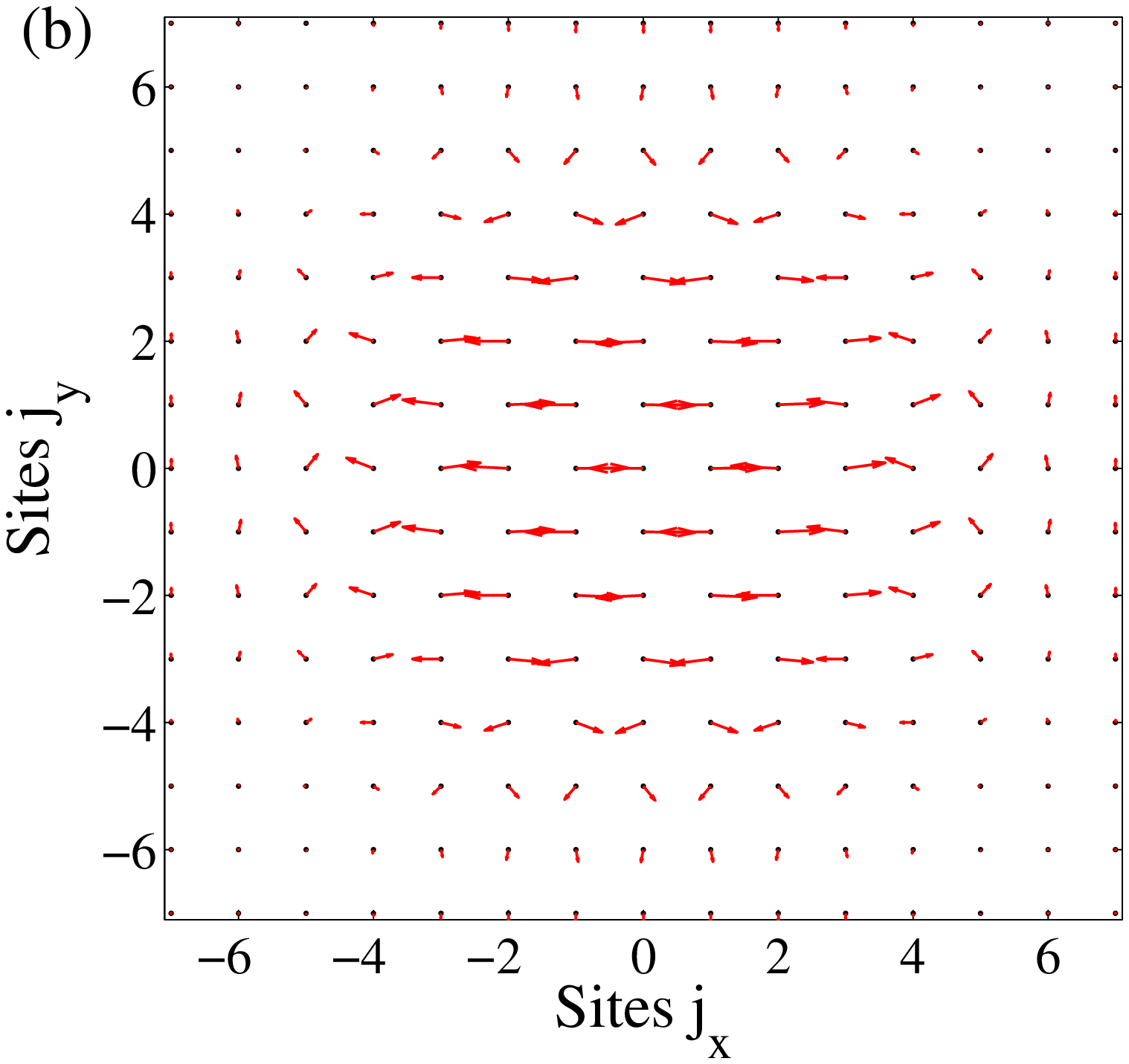}
\caption{(Color online) The Bloch vector at the different lattice
  sites (the $x$-component is strictly zero). The $y$-spin direction has been chosen along the horizontal axis and the $z$-spin direction along the vertical axis. The length of the vector
  represents the density, while the offset from the horizontal axis
  indicates breakdown of the anti-ferromagnetic order. The lattice sites are marked by black dots. The upper plot (a) gives the results where interaction plays a minor role, $U_0N=1$, while in (b) $U_0N=15$ and interaction cannot be ignored.   The rest of the parameters are the same as for Fig.~\ref{latticepop}.} 
\label{latticeBloch}
\end{figure}

\begin{figure}
\includegraphics[width=0.4\textwidth]{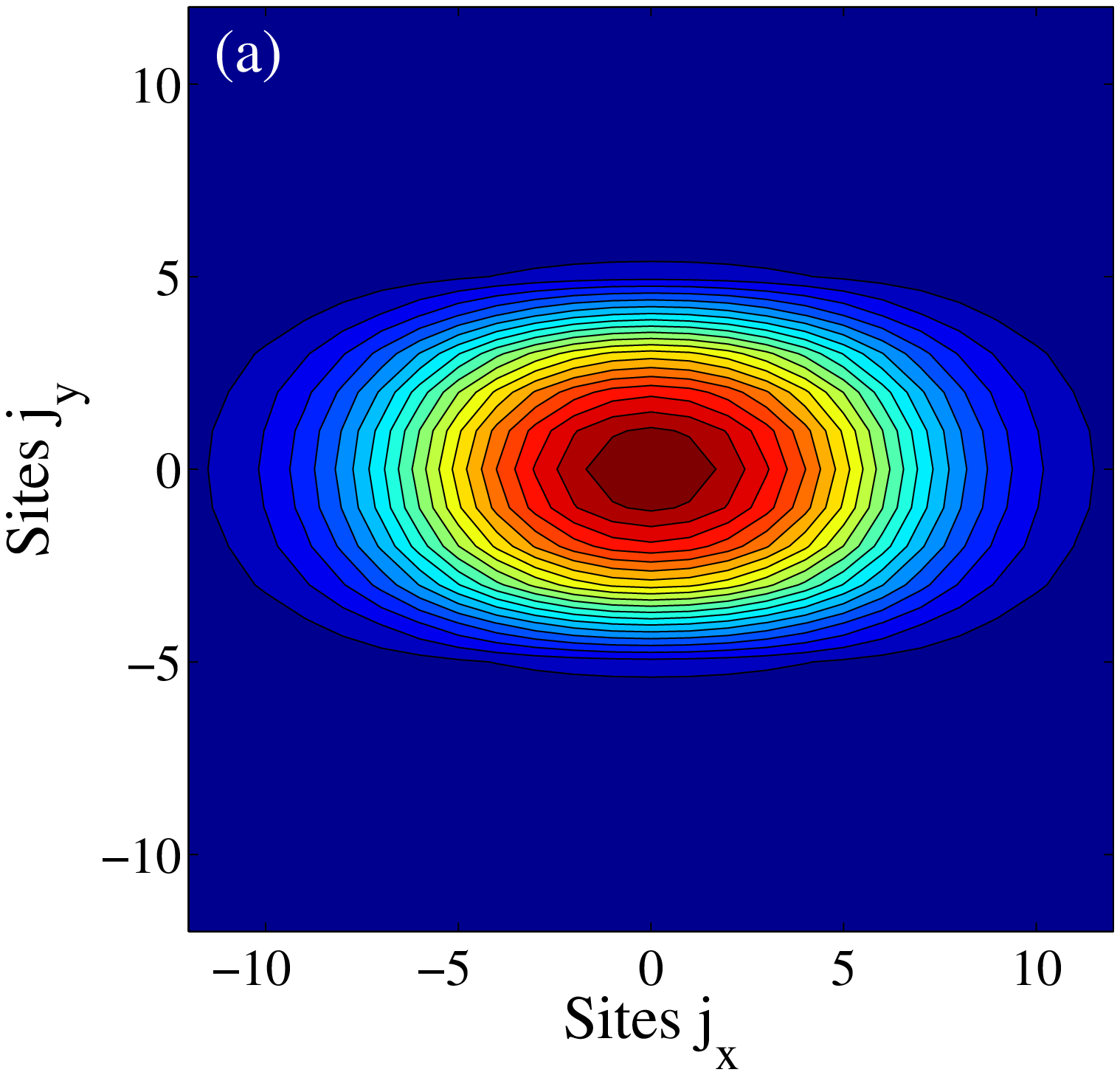}
\includegraphics[width=0.4\textwidth]{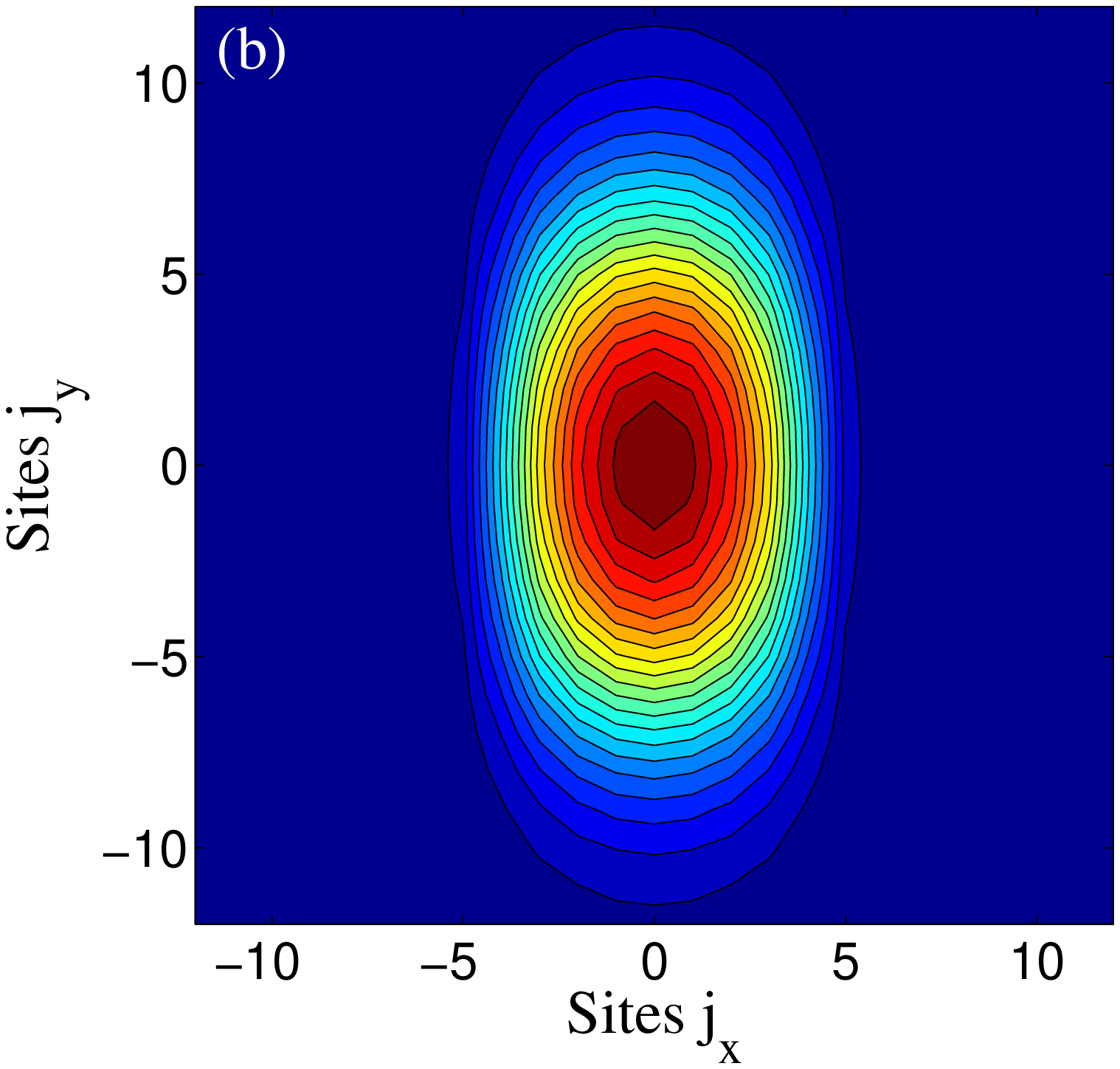}
\caption{(Color online) Plots showing the $x$- (a) and
  $y$-flavor (b) ground state populations, respectively, for the dimensionless system parameters $V_x=V_y=17$,
   $\omega=0.005$, and $U_0N=15$. Due to the larger interaction, the squeezing effect is not as pronounced in this case compared to Fig.~\ref{latticepop}.} 
\label{fig:unot15}
\end{figure}

The ground state lattice populations $|\psi_{x{\bf j}}|^2$ and
$|\psi_{y{\bf j}}|^2$ for a system of $V_x=V_y=17$, $\omega=0.005$,
and $U_0N=1$ are displayed in Fig.~\ref{latticepop} (a) and
(b). It is clear how the anisotropy manifest
itself, by rendering a condensate with spatially squeezed profile. 
In (c) we show the population imbalance $J_{z{\bf j}}$. 
As we argued above, whenever $J_{z{\bf
    j}}\neq0$ the anti-ferromagnetic order is broken, and from the
figure it is evident that this is especially true in the edge of the
condensate. To complement the results, we also present the
corresponding Bloch vectors in Fig.~\ref{latticeBloch} (a). Since
$J_{x{\bf j}}=0$, it is enough to show the Bloch vector in the spin $yz$-plane
${\bf J_j}=(0,J_{y{\bf j}},J_{z{\bf j}})$. By calling the horizontal axis the $y$-spin direction and the vertical axis the $z$-spin direction, we see that in the center of the condensate, the $J_{y{\bf j}}$
component dominates, while at the edge the Bloch vector no longer
points along the horizontal direction demonstrating the breakdown of
the anti-ferromagnetic order in these regions. Thus, at the center of the
condensate where $J_{z{\bf j}}\approx0$, the anti-ferromagnetic
ordering is still present.  

In Fig.~\ref{fig:unot15} we show the ground state lattice populations 
for a more strongly interacting system with $U_0N=15$. In this case
interactions and trap energies  are larger than the kinetic energy
and we approach the TF regime.
We can see how the effects of the anisotropic density are now smoothed and the
  region of the center of the trap is enlarged. The latter also
  corresponds to the region where non-trapped like physics actually
  occurs, as confirmed in Fig.\ref{latticeBloch} (b), by the
  presence of almost horizontal Bloch vectors. This also implies that
  now the ferromagnetic order extends over more sites in the lattice.

\begin{figure}
\includegraphics[width=0.5\textwidth]{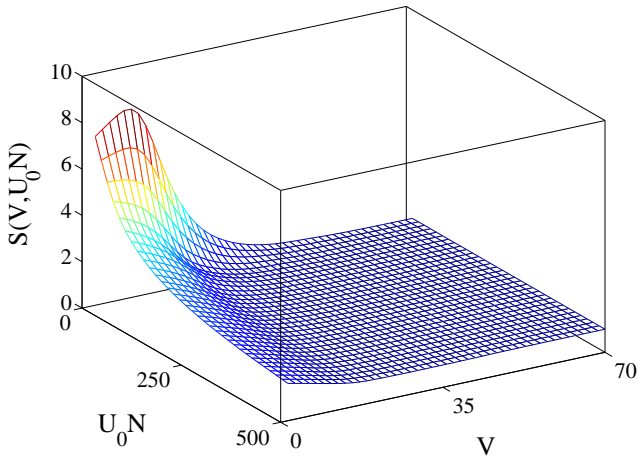}
\caption{(Color online) The condensate anisotropy parameter $S$ as defined in
  Eq.~(\ref{squeez}) as a function of the interaction strength $U_0N$
  and the lattice amplitude $V=V_x=V_y$. Whenever the amplitude or the
  interaction become large, the squeezing approaches one and the
  condensate enter into the TF regime. The dimensionless trap
  frequency $\omega=0.005$.} 
\label{squeezplot}
\end{figure} 

We complete the study of the interacting system's ground state in  
the symmetric lattice by investigating the behavior of the anisotropy
parameter (\ref{squeez}). Here, the relevant question to be understood
is related to characterization of $S$ when the system undergoes a
transition to the TF regime. When the kinetic energy
becomes suppressed, the anisotropy should vanish and hence
$S\rightarrow1$. In the lattice there are two ways of suppressing the
kinetic energy, either by increasing the interparticle interaction
strength $U_0$ directly by making use of Feshbach resonances, or by considering larger potential
amplitudes. 
The predicted behavior of $S$ is shown in
Fig.~\ref{squeezplot}. Note, that increasing $U_0N$ leads to a 
monotonic decrease of $S$ until it asymptotically reaches
1. In the other case, where variation of $V=V_x=V_y$ is considered,
$S$ also approaches 1 asymptotically, but now the behavior is not
monotonic. This anomalous and
surprising behavior does not appear in the continuum approximation. It
should be noted that the continuum limit is evaluated in the ideal
limit of $U_0=0$, and we especially have that $S_{con}$ is not
approaching 1 as $V\rightarrow\infty$. In this limit, on the other
hand, any small $U_0>0$ will imply $S=1$ since the kinetic term is
negligible compared to the interactions. For small and moderate $V$,
the continuum result (\ref{sqcon}) is found to increase monotonously
for increasing values of $V$. This behavior is not found for the
discrete model, even for $U_0=0$. Thus, for large amplitudes the
discrete and continuum models predict qualitatively different results
for the squeezed profile of condensate in terms of the
anisotropy parameter. We should also point out
that for small amplitudes, typically $V<5 E_r$~\cite{jonasNJP}, the
tight-binding approximations break down and the
results should not be taken too literally in this regime.   

\subsection{Properties in the anisotropic lattice}
Asymmetry in the lattice breaks the 
degeneracy of $x$ and $y$ flavors.  
In order to investigate the effect of anisotropies we introduce
the asymmetry parameter 
\begin{equation}
R=\frac{V_x}{V_y}
\end{equation}
which controls the ratio between the lattice depths, such that $R=1$
represents the symmetric lattice configuration we discussed earlier. 
We have numerically verified that the dominant effect of the asymmetry is to shift the energy levels of $x$- and $y$-flavors. By considering only a single site first, we note that in the harmonic approximation this shift equals
\beq
\Delta=E_{y}-E_{x}=2\sqrt{V_x}\left(\sqrt{R}-1\right),
\enq
where $E_x$ and $E_y$ are the energies of the onsite flavors, i.e. $E_\alpha=\int\,d\vec{r}w_{\alpha{\bf j}}^*(\vec{j})\left[-\nabla+V_{latt}(\vec{r})\right]w_{\alpha{\bf j}}(\vec{j})$ and where the ${\bf j}$ dependence vanishes. In this single site picture, this splitting will have only a small effect if it is much smaller than the characteristic interaction energy scale $E_{int}\sim U_0N |\psi_x|^2$. 

The picture becomes more complicated when we consider more sites. 
It can be, for example, that the region $\delta$ around $R=1$ 
in which interaction mixes the two flavors changes as the trapping strength
is varied, and in particular, if $\delta$ is small, the properties of
the ground state may change dramatically with small variations in the various lattice parameters. On the other hand, if these parameters can be controlled, the physics around the degeneracy point might lead to novel physics similar to the
adiabatic driving considered recently in Ref.~\cite{hemmerich3}. However, it is worth pointing out that the present model possesses an additional property, namely that the $x$- and $y$-flavor densities are spatially different and adiabatic driving
between the two might therefore lead to macroscopic particle flow
within the trap. While interesting, this time-dependent aspect will be
addressed elsewhere. 

The asymmetries for our square lattice
can in principle be implemented in two ways, either by considering a lattice with
different wave vectors $k_x$ and $k_y$ or different amplitudes $V_x$
and $V_y$. Here we characterize the behavior of the system in the
latter process. 
\begin{figure}
\includegraphics[width=0.5\textwidth]{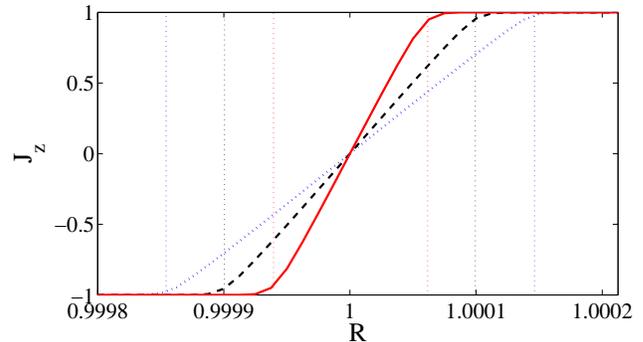}
\caption{(Color online) The parameter $J_z$ as a function of the
  lattice asymmetry parameter $R$, for three different trapping
  frequencies, $\omega=0.003$ (red solid line), $\omega=0.005$ (black
  dashed line), and $\omega=0.007$ (blue dotted line). The vertical dashed thin lines indicate the typical sizes of $\delta$ which determines the transition region where the two atomic flavors coexist. It is clear how $\delta$ is decreased when the trap is ``opened up'' (decreasing $\omega$). The remaining dimensionless parameters are $U_0N=1$ and
  $V_x=17$ (meaning that $V_y=17/R$).}  
\label{fig:Rplot}
\end{figure}
The sensitivity to $R$
can be analyzed, for example, in the value of the mean population inversion 
\begin{equation}\label{meanJ}
J_{z}=\frac{1}{N}\sum_{\bf j}J_{z{\bf j}}.
\end{equation}
If $J_z=-1$, the system consists of only $y$-flavor atoms, and
$J_z=+1$ represents only atoms in the $x$-flavor. Thus, $J_z$ gives a measure of how much interaction mixes the two flavors. In the vicinity of $R = 1$,
the properties of  $J_z$ are illustrated in Fig. (\ref{fig:Rplot}). It clearly shows uniquely occupied flavors in both regions where $R < 1$ and $R > 1$. Also, as expected, the exact point $R = 1$ is characterized by equal sharing of population
among the two flavors, and therefore one recovers the properties of the
degenerate system. As pointed out above, the non-zero interaction ($U_0\neq0$) is
crucial in order to stabilize the equal population at $R=1$. 

The confinement imposed by the harmonic trap implies that we are
dealing with a finite size system. The frequency $\omega$ sets, in
some sense, the system size and, as we discussed above, it is interesting to understand how $\delta$ depends on the system size. Figure (\ref{fig:Rplot})
depicts the variations of $J_z$ around $R=1$, and it is seen how these
become more dramatic when $\omega$ is decreased. More precisely, there
seems to be a one-to-one correspondence between the range $\delta$ in
which $|J_{z}|<1$ and $\omega$, and as $\omega\rightarrow0$ the plot
indicates also that $\delta\rightarrow0$. 
This suggests (for very weakly interacting systems) similar
behavior to the one generally exhibited by systems undergoing a
first order phase transition~\cite{sachdev}. In addition, we also
studied the ground state energy $E_0(R)$ and found that $dE_0(R)/dR$
shows a pronounced change around $R=1$ as $\omega$ is decreased. We have also numerically verified that the range $\delta$ grows for increasing interaction strength $U_0N$ in agreement with our earlier argument that interaction mixes the flavors.

The above findings suggest that for weak interactions 
a careful adjustment of the lattice is
required in order to study the anti-ferromagnetic properties. 
As interactions become stronger the anti-ferromagnetic properties
become more robust.
In experimental realizations even a small
temperature might actually help to establish a phase
coherence between $x$- and $y$-flavor atoms since the energy gap between the ground and first excited energies greatly decreases around the $R=1$ point and in its vicinity one may expect population also of the first excited state. We furthermore notice that for non-zero $\omega$, the transition from one to the other extreme of $J_z$ is smooth, and therefore by controlling the lattice amplitudes the system could be considered for studies of the many-body Landau-Zener transition~\cite{altland} or the Kibble-Zurek mechanism~\cite{kz}.

\section{Conclusion}
We have investigated how a confining potential affects the properties of
bosonic atoms residing on the $p$-bands of optical lattices. Our focus
was on the 2D square lattice with equal lattice amplitudes in the two
directions and we restricted our analysis to a mean-field
approach. It is known that for a $p$-band square lattice model, even at a
mean-field level the ground state forms non-trivial states in terms of
an anti-ferromagnetic order~\cite{isaksson,ph4}. As a result of the anisotropic tunneling on the
$p$-band together with the confinement introduced by the trap, we 
showed that the anti-ferromagnetic structure is destroyed in the 
edges of the condensate. The effects of the tunneling anisotropy are
also manifest in the density profile of the atomic cloud, yielding 
a spatially elongated condensate in one of the two spatial directions,
despite the isotropic trap. We showed how this narrowing is suppressed when the kinetic
energy is lowered, either due to increasing of the strength of
atom-atom interactions and/or due to increasing the lattice amplitudes. The same suppression was found also for the ideal gas when
the temperature is increased and thereby the properties of the gas are
greatly determined by thermal atoms. By considering unequal lattice
amplitudes in the $x$- and $y$-directions, the degeneracy on the
$p$-bands is broken, and we demonstrated that the sensitivity of the ground state properties depend strongly on the system ``size". The results presented are for 2D lattices, but it
is understood that the general findings directly generalize to 3D as
well. In the 3D cubic case, the phase ordering can be more 
complicated~\cite{ph4},
but as in the 2D case, this ordering would also be destroyed in the
edges of the condensate in a trapped system.

One point we have not addressed concerns experimental
realizations. The main source for dissipation and decoherence in the
square lattices is scattering of two $p$-band atoms into one $s$- and
one $d$-band atom~\cite{ph4,pexp2}. This process is resonant in the
harmonic approximation, while it is generally off-resonant for actual lattices,
which causes the typical life-time for $p$-band atoms to be much
larger than the characteristic tunneling times. In Ref.~\cite{pexp2},
coherence of $p$-band atoms in a cubic lattice was indeed
demonstrated. Alternatives for suppressing this decay further include
loading fermionic atoms into the $s$-band of the lattice~\cite{sWu} or
considering experimental setups with non-separable
lattices~\cite{hemmerich1,hemmerich2,fernandafriend}. In 
the first case, the presence of fermions in the $s$-band prevents the
bosonic $p$- band atoms to occupy the lowest band due to atom-atom
interactions. Now in configurations involving non-separable lattices
(e. g. superlattices), few bands can be separated from the
rest, and thus the role of the ($p + p \rightarrow s + d$)
scattering becomes overshadowed. In Refs.~\cite{hemmerich1,hemmerich2}, however, the
experimental setup gives rise to hybridization of different flavor atoms and
the analysis becomes more complex than the one for the simple square
lattice considered here.   

Another important experimentally relevant question concerns detection of the
presented predictions. If the detection makes no difference between $x$- and
$y$-flavor atoms, the Bloch vector cannot be fully measured. However,
in a recent work it was suggested how such measurements can indeed be
performed~\cite{meas}. The idea utilizes Raman pulses that rotate the
spinor (\ref{spinorder}) similar to qubit measurements in atomic
physics~\cite{haroche}. Moreover, in a recent experiment on triangular
lattices~\cite{sengstock} it was demonstrated how the phase of the
condensate affects the densities in time-of-flight measurements. We
have numerically studied the full condensate order parameter
$\Psi(x,y)$, and found that coherence within single sites are seen in
$\Psi(x,y)$ while long range coherence is manifested in the momentum
distribution of $\Psi(x,y)$. This means that if the condensate
density $|\Psi(x,y)|^2$ is detected at different time instants
in a time-of-flight measurements, one could in principle extract all
information about the phase coherence. 

We believe that entering the more strongly correlated regime where
quantum fluctuations become more important would be of interest. The
mean-field method adopted here is not capable of capturing these
effects, and we therefore leave this investigation for the future. We
especially intend to study the ``wedding cake''  
structure~\cite{wedding} formed by
alternating insulating Motts and superfluids in the presence of  a
harmonic trap, as well as non-equilibrium properties of the system.

\begin{acknowledgments}
Financial support from the Swedish Research Council (Vetenskapsr\aa det) is acknowledged. JL acknowledges financial support from DAAD (Deutscher Akademischer Austausch Dienst) and the Royal Research Council Sweden (KVA). JPM acknowledges financial support from the Academy of Finland (Project 135646).
\end{acknowledgments}

\end{document}